\documentclass[12pt]{article}
\usepackage{amsfonts}
\usepackage{bm}
\usepackage{amsmath}
\usepackage{epsfig}

\usepackage{amssymb,lscape}
\usepackage[matrix,arrow,curve]{xy}

\newcommand{\bmat}{\left(\begin{array}}
\newcommand{\emat}{\end{array}\right)}

\def\gtrsim{\mathrel{\raise.3ex\hbox{$>$\kern-.75em\lower1ex\hbox{$\sim$}}}}

\def\-{\hphantom{-}}

\def\s2{\frac{1}{\sqrt2}}

\def\mg{m_{3/2}}
\def\mg2{m^2_{3/2}}

\def\Dsl{\,\raise.15ex\hbox{/}\mkern-13.5mu D} %this one can be subscripted

\def\be{\begin{equation}}
\def\ee{\end{equation}}
\def\bea{\begin{eqnarray}}
\def\eea{\end{eqnarray}}

\newcommand{\nn}{\nonumber}

\begin{document}

\pagestyle{plain}

%----------------------------------------------------------------------%
%  numbering equations with section number
%----------------------------------------------------------------------%
\makeatletter
\@addtoreset{equation}{section}
\makeatother
\renewcommand{\theequation}{\thesection.\arabic{equation}}
%----------------------------------------------------------------------%
%  title page
%----------------------------------------------------------------------%
\pagestyle{empty} 
%\rightline{IPhT-???/???}
\begin{center}
\LARGE{The effective action of Double Field Theory\\[10mm]}
\large{ Gerardo Aldazabal${}^{a,b}$, Walter Baron${}^c$,  Diego Marqu\'es$^d$ \\
and  Carmen
N\'u\~nez$^{c,e}$
 \\[6mm]}
\small{
${}^a${\em Centro At\'omico Bariloche, ${}^b$Instituto Balseiro
(CNEA-UNC) and CONICET} \\[-0.3em]
8400 S.C. de Bariloche, Argentina.}\\
${}^c$ {\em Instituto de Astronom\'ia y F\'isica del Espacio
(CONICET-UBA)} \\
C.C. 67 - Suc. 28, 1428 Buenos Aires, Argentina.  \\
%[1cm]
${}^d${\it  Institut de Physique Th\'eorique,
CEA/ Saclay}  , \
91191 Gif-sur-Yvette Cedex, France.\\
${}^e${\em Departamento de F\' isica, FCEN, Universidad de Buenos Aires}\\
[1cm]

\small{\bf Abstract} \\[0.5cm]\end{center}
We perform a generalized Scherk-Schwarz dimensional reduction of
Double Field Theory on a twisted double torus. The four dimensional
effective action is shown to exactly reproduce the bosonic electric sector
of gauged ${\cal N} = 4$ supergravity. We present explicit
expressions for the gaugings in terms of the twists, and analyze the
associated string backgrounds. This framework provides a higher 
dimensional origin of the gaugings in terms of generalized fluxes.

{\small}

\newpage
%----------------------------------------------------------------------%
%  Resetting of counters
%----------------------------------------------------------------------%
\setcounter{page}{1}
\pagestyle{plain}
\renewcommand{\thefootnote}{\arabic{footnote}}
\setcounter{footnote}{0}
%----------------------------------------------------------------------%
%  Paper begins
%----------------------------------------------------------------------%

\tableofcontents

\section{Introduction}

Flux compactifications of string theory \cite{Grana:2005jc} lead to gauged
supergravities \cite{Samtleben:2008pe},
 providing an efficient mechanism of moduli stabilization and spontaneous
supersymmetry breaking.
An intriguing puzzle is that gauged supergravity contains more gaugings
than those that can be reached through geometric compactifications of the
different string
effective actions or
string supergravities.
 The presence of additional (non-geometric) gaugings suggests that some
features of
 string compactifications have  not yet been properly taken into account.

\begin{figure}[h]
\centerline{\includegraphics[width=16 cm]{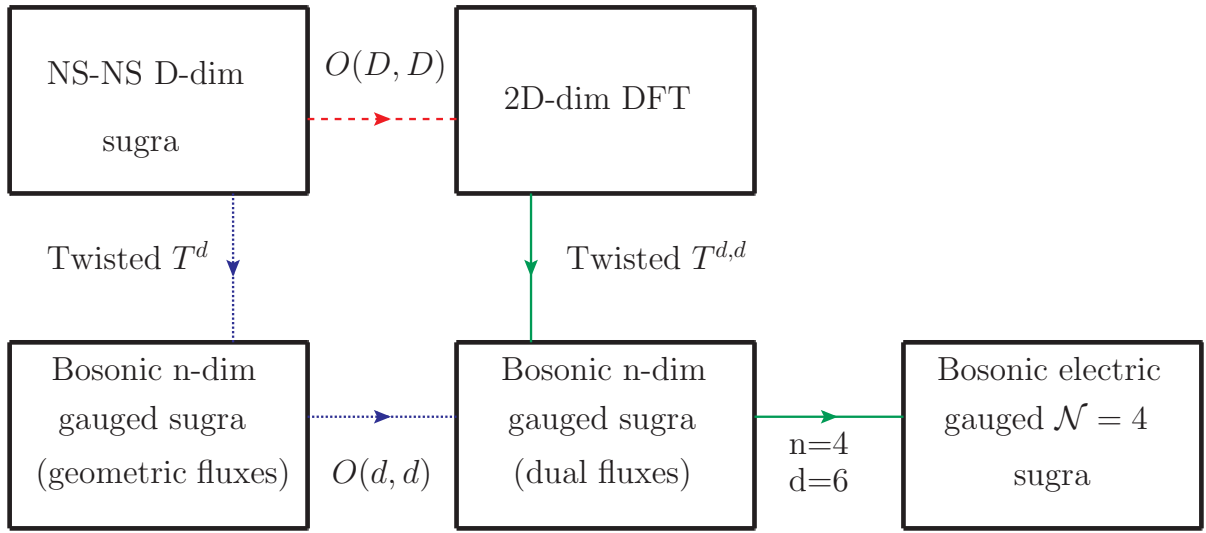}} \label{fig}
\caption{From string supergravity to four dimensional ${\cal N}=4$ 
 gauged 
supergravity.}
\end{figure}

The missing gaugings could be obtained
 by invoking U-duality arguments at the level of the
dimensionally reduced effective actions
\cite{Shelton:2005cf}-\cite{Aldazabal:2010ef}. This approach
corresponds to the dotted (blue) path in Figure \ref{fig}.
Although efficient in generating the full set of gaugings, this
procedure presents the disadvantage of leaving many of them
unexplained in terms of a higher dimensional origin. For instance,
enforcing T-duality invariance requires the introduction of
non-geometric fluxes. This can be seen by comparing the effective
four dimensional superpotentials arising in orientifold
compactifications of type IIA and IIB  ten dimensional supergravity
actions,
 where a broader class of internal spaces is needed to have a geometric
interpretation.

The concepts of Generalized Complex Geometry
\cite{Hitchin:2004ut}-\cite{Grana:2008yw}, non-geometry
\cite{Hellerman:2002ax,Dabholkar:2002sy} and double geometry
\cite{Hull:2004in}-\cite{Hull:2009sg}   were proposed as  suitable
frameworks to deal with this situation
(see also \cite{b2, b3} for extensions to M-theory). 
By treating the symmetries
of the NSNS antisymmetric tensor field and diffeomorphisms on an
equal footing, some of the properties of T-duality are naturally
incorporated. Inspired by this approach, a way to obtain the usual
contribution of the locally geometric $Q$-flux \cite{Kachru:2002sk}
to the four-dimensional effective action, from a ten dimensional one, 
was recently proposed in \cite{Andriot:2011uh}.

On the other hand, there have also been attempts to promote U-duality to a
symmetry at the level of the higher
dimensional supergravity.
 For T-duality
 this was successfully achieved by Double Field Theory (DFT), originally
formulated in \cite{Hull:2009mi}-\cite{Hohm:2010pp} and extended in
\cite{Hohm:2011cp}-\cite{Kwak:2010ew}. This attempt is represented
in the figure with the dashed (red) arrow. The idea, which we review
in Section \ref{ReviewDFT}, was to introduce additional coordinates
to the standard space-time embedding of closed strings in toroidal
backgrounds, dual to winding. In DFT the fields  depend on both sets
of
 coordinates and this gives rise to a
$2D$-dimensional theory formulated on a double space. The stringy nature
of the theory is manifested in the
fact that DFT is T-duality invariant, so it promotes a string duality to
a symmetry.

In this work we follow the path indicated by the two solid (green)
arrows in Figure \ref{fig}. First, by performing a generalized
Scherk-Schwarz dimensional reduction \cite{Scherk:1979zr} on a
twisted double torus we obtain the effective action of DFT. This
procedure allows to identify the new degrees of freedom present in
DFT as the origin of the missing (non-geometric) gaugings. The
computation,  for an arbitrary number of dimensions, is performed in
Section \ref{SSDR}. The reduced theory is a generalization of that
found in \cite{Kaloper:1999yr}. Interestingly enough,
 it
 includes geometric fluxes as well as the locally (but not globally)  geometric
fluxes $Q$ and the  locally non-geometric fluxes $R$.
 Besides, it inherits a remnant
gauge symmetry from DFT based on the C-bracket, and a (would be)
global $O(d,d)$ invariance which is broken by the fluxes.

 In Section \ref{gaugedsugra} we show that the four dimensional
 effective action with six internal dimensions is dual to the bosonic
electric sector of gauged ${\cal N} = 4$ supergravity, as formulated
in \cite{Schon:2006kz}. This corresponds to the second solid (green)
arrow. In order to recover the full set of gaugings, we show how to
generalize the Scherk-Schwarz reduction to include the fundamental
fluxes $\xi_{+M}$. Additionally, we enhance the $O(D,D)$ symmetry of
DFT to $O(D,D+N)$ by adding $N$ generalized vector fields along the
lines of \cite{Andriot:2011iw} and \cite{Hohm:2011ex}.

 Section \ref{NonGeomSec} is devoted to the analysis of the string
(non)geometric backgrounds and
the fluxes they give rise to. We show that the internal space can be
thought of as a twisted double torus.
 Alternative interpretations of our results from the
point of view of superstring compactifications are discussed. In
particular we show that the resulting effective theory can be seen
either as a bosonic sector of heterotic string compactification,
extended by T-duality \cite{Shelton:2005cf, Aldazabal:2010ef},  or
as a bosonic sector of a Type IIB orientifold compactification
\cite{Aldazabal:2008zza, Aldazabal:2011yz}.

Finally, the conclusions in Section \ref{conclu} contain a summary
of our results. We include an Appendix illustrating how single-flux string
backgrounds are encoded in the
twisted double torus.

\section{Review of Double Field Theory}\label{ReviewDFT}

$D$-dimensional closed strings
on toroidal backgrounds carry both momentum and winding, the former being dual
to space-time coordinates.
Double Field Theory  was constructed out of the idea of assigning dual
coordinates also to winding. While in string
%the NS-NS sectors of string
supergravities in $D$ dimensions
%are functionals of
fields depend  only on space-time coordinates, DFT incorporates a
%generalizes the bosonic
%NS-NS sector to include
dependence also
on the coordinates dual to winding. For this reason, it is a theory defined in a
doubled space with coordinates
$X^{\hat M} = (X^{\hat \mu},\tilde X_{\hat \mu})$.
The most remarkable feature of this theory is that it is invariant
under T-dualities, and more generally under the full
$O(D,D)$ group associated to the isometries of the doubled torus. In this way,
DFT is a field theory that takes ``stringy''  features into
account  by promoting a string duality to a symmetry.

In this section we present a brief review of the generalized metric formulation
of DFT,
mainly to exhibit the results that we will use throughout the paper.
%The notation is detailed in the
%Appendix.
The notation basically  coincides with the standard conventions, but  we put a
hat on all
fields and indices
to facilitate the reading of the forthcoming sections.

The building block of the theory is the generalized $2D \times 2D$ metric
\bea
\hat{\cal H}_{\hat M\hat N}
                =\left(\begin{matrix}{\hat g}^{\hat \mu\hat \nu}
&    -{\hat g}^{\hat \mu\hat \rho}{\hat B}_{\hat \rho\hat \nu}        \cr
                                  {\hat B}_{\hat \mu\hat \rho}{\hat g}^{\hat
\rho\hat \nu}
&   {\hat g}_{\hat \mu\hat \nu}-{\hat B_{\hat \mu\hat \rho}{\hat g}^{\hat
\rho\hat \sigma}
{\hat  B}_{\hat \sigma\hat \nu}}  \end{matrix}\right)\, ,\label{MetricaGen}
\eea
constructed out of the $D$-dimensional metric $\hat g_{\hat \mu \hat \nu}
(X^{\hat \mu},
\tilde X_{\hat \mu})$ and the $D$-dimensional $2$-form
field $\hat B_{\hat \mu \hat
\nu} (X^{\hat \mu},\tilde X_{\hat \mu})$.
This generalized metric is an element of $O(D,D)$, so it preserves
the metric of
the group that we take
of the following form
\begin{equation}
\hat {\cal I} = \left(\begin{matrix}0& 1_D\\ 1_D & 0\end{matrix}\right)\,
.\label{MetricaDD}
\end{equation}
In addition, the model also contains an $O(D,D)$ invariant dilaton
\be
e^{-2 \hat d} = \sqrt{|\hat g|} e^{-2\hat \phi}\, .\label{Dilaton}
\ee
The background independent formulation of DFT is defined by an
action that can be written in a
compact form (up to total derivatives) in terms of a generalized Ricci scalar as
\bea
S_{\rm DFT}=\int d^{D}X d^D {\tilde X} \ e^{-2\hat d}\ {\cal R}
({\hat{\cal H}},\hat d)\, ,\label{action}
\eea
where ${\cal R}$ is defined  by
\bea
{\cal R}({\hat{\cal H}},\hat d)&=&4{\hat{\cal H}}^{\hat M\hat N}\partial_{\hat
M}\partial_{\hat N} \hat
d-\partial_{\hat M}\partial_{\hat N}\hat{\cal H}^{\hat M\hat N}-4\hat{\cal
H}^{\hat M\hat N}\partial_{\hat M} \hat
d\partial_{\hat N} \hat d +
 4\partial_{\hat M} \hat{\cal H}^{\hat M\hat N}\partial_{\hat N} \hat d \cr
&&+\ \frac18 \hat{\cal H}^{\hat M\hat N}\partial_{\hat M}\hat{\cal H}^{\hat
K\hat L}\partial_{\hat N}
\hat{\cal H}_{\hat K\hat L}-\frac12\hat{\cal H}^{\hat M\hat N}\partial_{\hat
M}\hat{\cal H}^{\hat K\hat L}
\partial_{\hat K} \hat{\cal H}_{\hat N\hat L}\, .
\eea
When the coordinates dual to winding are frozen, $i.e.$ $\partial_{\tilde X}~
{\bullet} = 0$, this action
 reduces
to the standard NS-NS bosonic sector of the $D$-dimensional string
supergravities. The theory is
constrained by  a world-sheet level-matching condition that  can be written
as
\be
\partial_{\hat M}\partial^{\hat M} ~ {\rm\bullet}= 0 \ ,
\label{strong constraint} \ee and the construction of the action
(\ref{action}) requires the so called ``strong constraint'' stating
that all fields and their products (represented by $\bullet $ above)
must be annihilated by the differential operator   (\ref{strong
constraint}). The strong constraint implies that locally there is
always an $O(D,D)$ transformation that rotates into a frame in which
fields depend only on half of the coordinates.

Regarding symmetries, the action is manifestly invariant under global  $O(D,D)$
transformations acting as
\be
{\hat{\cal H}}_{\hat M\hat N} \to \hat U^{\hat A}{}_{\hat M} \hat {\cal H}_{\hat
A \hat B} \hat U^{\hat B}{}_{\hat N} \ , \ \ \ \ \ \  \hat d \to \hat d \ , \ \
\ \ \ \ \hat U \in O(D,D)\, .\label{ODDinv}
\ee
In addition it has the following gauge symmetry
\bea
\delta_{\hat\xi}  \hat d &=& \hat \xi^{\hat M}\partial_{\hat M} \hat d -
\frac 12 \partial_{\hat M} \hat \xi^{\hat M} \, ,\nn \\
\delta_{\hat\xi} \hat{\cal H}^{\hat M\hat N}&=&\hat\xi^P\partial_P \hat{\cal
H}^{\hat M \hat N}
                  +\left(\partial^{\hat M}\hat\xi_{\hat P}-\partial_{\hat P}
\hat\xi^{\hat M} \right)
\hat{\cal H}^{\hat P\hat N}
                  +\left(\partial^{\hat N}\hat\xi_{\hat P}-\partial_{\hat P}
\hat\xi^{\hat N} \right)
\hat{\cal H}^{\hat M\hat P}\, .\label{gauge2D}
\eea
This transformation rule is an $O(D,D)$ covariant extension of the standard Lie
derivative that governs infinitesimal
diffeomorphisms. Generalized Lie derivatives acting on vectors can be
constructed as
\be
\hat {\cal L}_{\hat \xi}  A_{\hat M} = \hat \xi^{\hat N} \partial_{\hat N}
A_{\hat M} + (\partial_{\hat M}
\hat\xi^{\hat N} - \partial^{\hat N} \hat\xi_{\hat M})  A_{\hat N}\,
,\label{c-LD}
\ee
such that the transformation rules (\ref{gauge2D}) read
$\delta_{\hat \xi} \hat {\cal H}_{\hat M \hat N} = \hat {\cal L}_{\hat \xi} \hat
{\cal H}_{\hat M \hat N}$.
The gauge transformations then  close under the so-called C-bracket, defined as
\be
\left[\hat {\cal L}_{\hat \xi_1},  \hat {\cal L}_{\hat \xi_2} \right] A_{\hat M}
= -
\hat {\cal L}_{[\hat \xi_1, \hat\xi_2]_{\rm C}} A_{\hat M}\, ,
\ee
where
\be
\left[\hat\xi_1, \hat \xi_2\right]^{\hat M}_{\rm C} = 2 \hat \xi_{[1}^{\hat N}
\partial_{\hat N}
\hat \xi^{\hat M}_{2]} -  \hat\xi^{\hat P}_{[1} \partial^{\hat M} \hat
\xi_{2]\hat P}\, .\label{Cbracket}
\ee

For further insight on DFT  we refer to the original works
 \cite{Hull:2009mi}-\cite{Hohm:2010pp}
and to their many extensions \cite{Hohm:2011cp}-\cite{Kwak:2010ew}.

\section{Scherk-Schwarz dimensional reduction}\label{SSDR}

In this section we perform the twisted dimensional reduction of DFT
and compute the effective action. %Notations and conventions for
%indices are specified in the Appendix.
The starting point for the
compactification procedure
is the generalized $2D$-dimensional double space on which DFT is
defined. We identify $n$ of its coordinates with space-time
coordinates and other $n$ with their duals. The first step is to
compactify this double space-time on a double torus of vanishing 
dual radius, so
that the usual space-time is naturally decompactified. We are therefore
effectively left with a $(n + 2d)-$dimensional space ($D = n + d$)
that we compactify on a twisted $T^{d,d}$ torus to obtain the
$n$-dimensional effective theory of DFT.

Throughout the paper we use many different indices. The original
formulation of DFT in $2D$-dimensions is based on the usual $D$
 ``string'' coordinates and the remaining $D$ correspond
 to the dual space.
After dimensionally reducing in a $2d$-dimensional space, the effective action
is an $n = D-d$ dimensional theory (provided the coordinates dual to
 space-time are taken to vanish).
Of course, at the end of the day our main interest is in $D = 10$, $d = 6$
and $n = 3+1$, but since most
computations can be performed without explicitly specifying the dimensions,
we keep the results as general as possible.
We use the following notation:

{\it Curved} and {\it tangent}
 indices are respectively
\bea
\hat M, \hat N, \hat O, \hat P, \dots \in\{0,1,2,\dots,2D-1 \} \ ,
 \ \ \  \hat A, \hat B, \hat C, \hat D, \dots
\in\{0,1,2,\dots,2D-1 \}\, ,
\eea
for
 the full $2D$ space;
\bea
\hat \mu, \hat \nu, \hat \xi, \hat \rho, \hat \sigma, \dots \in\{0,1,2,\dots,D-1
\}
\, , \ \ \ \hat m, \hat n, \hat o, \hat p, \hat q, \dots \in\{0,1,2,\dots,D-1
\},
\eea
for the $D$-dimensional ``stringy''  coordinates;
\bea
M,N,O,P,\dots\in\{1,2,\dots,2d\}
\, , \ \ \ A,B,C,D,\dots\in\{1,2,\dots,2d\},
\eea
for $2d$-dimensional internal and dual coordinates (which we denote by ${\mathbb
Y}^M$);
\bea
\mu, \nu,  \rho, \sigma,\dots  \in\{0,1,\dots , n-1 \} \, , \ \
\ m,n,p,q,\dots\in\{0,1,\dots , n-1 \}, 
\eea 
for spacetime and 
\bea \alpha, \beta,
\gamma, \delta,\dots  \in\{1,2,\dots , d \} \, , \ \ \
a,b,c,d,\dots\in\{1,2,\dots , d \}, 
\eea
for the $d$-dimensional
internal indices (with the corresponding coordinates denoted as
$y^\alpha$). Dual coordinates are $\tilde y_\alpha$ and 
we usually write ${\mathbb Y}^A=( \tilde  y_a, y^a)$.

As we have seen, the degrees of freedom in DFT are represented by an
invariant dilaton $\hat d$ defined in (\ref{Dilaton}) and a  generalized
$2D\times 2D$ metric (\ref{MetricaGen}), namely
\bea
\hat{\cal H}_{\hat M\hat N}=\left(\begin{matrix}\hat{\cal H}^{\hat \mu\hat \nu}
&    \hat{\cal H}^{\hat \mu}{}_{\hat \nu}                         \cr
                                  \hat{\cal H}_{\hat \mu}{}^{\hat \nu}
&    \hat{\cal H}_{\hat \mu\hat \nu}
\end{matrix}\right)
                =\left(\begin{matrix}{\hat g}^{\hat \mu\hat \nu}
&    -{\hat g}^{\hat \mu\hat \rho}{\hat B}_{\hat \rho\hat \nu}        \cr
                                  {\hat B}_{\hat \mu\hat \rho}{\hat g}^{\hat
\rho\hat \nu}
&   {\hat g}_{\hat \mu\hat \nu}-{\hat B_{\hat \mu\hat \rho}{\hat g}^{\hat
\rho\hat \sigma}
{\hat  B}_{\hat \sigma\hat \nu}}  \end{matrix}\right)\,
.\label{MetricaGeneralizada}
\eea
This metric can be constructed out of a vielbein with triangular\footnote{We
refer to the Appendix
for a discussion on the generality of considering the triangular gauge.}
form, $i.e.$
\bea
\hat {\cal E}=\left(\begin{matrix} (\hat e^{-1})^T     &   -(\hat e^{-1})^T \hat
B       \cr
                                          0            &            \hat e
\end{matrix}\right)\, ,\label{TriangularGauge}
\eea
which is related to  (\ref{MetricaGeneralizada}) through
\bea
\hat {\cal H}=\hat {\cal E}^T\,\hat \eta \,\hat {\cal E}\ , \ \ \ \ \ \ \ \ \ \
\hat \eta=\left(\begin{matrix} \eta^{-1}     &      0        \cr
                                    0        &     \eta
\end{matrix}\right)\, ,
\eea
where $\eta = diag(-+\cdots +)$.

We have defined the $D$-dimensional fields that
parameterize the generalized metric, namely the Kalb-Ramond field
$\hat B$ and the $D$-bein $\hat e$ associated to the metric
\be\hat g_{\hat \mu\hat \nu}=\hat e^{\hat m}{}_{\hat \mu}\hat\eta_{\hat m\hat
n}\hat e^{\hat n}{}_{\hat \nu}=\left(\hat e^T\hat \eta \hat
e\right)_{\hat \mu\hat \nu} \, .\ee

These  can then be written as usual in terms of the compactified
degrees of freedom as 
\bea 
\hat g_{\hat \mu \hat \nu}=
\left(\begin{matrix}
                                                 g_{\mu\nu}+
h_{\gamma\delta} {\cal A}^\gamma{}_\mu {\cal A}^\delta{}_\nu   &
                          {\cal A}^\gamma{}_\mu h_{\gamma\beta}
     \cr
     h_{\alpha\gamma}{\cal A}^\gamma{}_\nu         &
                                                 h_{\alpha\beta}
                             \end{matrix}\right)
\ , \ \ \ \ \ \hat B_{\hat \mu \hat \nu}=  \left(\begin{matrix}
          \breve{B}_{\mu\nu}                                      &
              -\breve{\cal{B}}_{\beta}{}_{\mu}         \cr
   \breve{\cal{B}}_{\alpha}{}_{\nu}                      &
      b_{\alpha\beta}           \end{matrix}\right)\, .\label{VectoresDesde}
\eea

From the $n$-dimensional point of view, the $D$-dimensional metric
decomposes into an $n$-dimensional metric $g_{\mu\nu}$, $d$ vectors
${\cal A}^\alpha_{\ \mu}$ and $d(d+1)/2$ real scalars
$h_{\alpha\beta}$. The $B$-field also decomposes into an
$n$-dimensional $2$-form $B_{\mu\nu}$,  $d$ vectors ${\cal
B}_{\alpha \mu}$ (which we will further relate to $\breve{
B}_{\mu\nu}$ and $\breve{\cal B}_{\alpha \nu}$ in equation
(\ref{B})) and $d(d-1)/2$ real scalars $b_{\alpha\beta}$.

Having set the structure of the (compactified) fields, we proceed
with the Scherk-Schwarz twist. The global symmetry of the action
that we intend to twist  is the $O(D,D)$ invariance defined in
(\ref{ODDinv}). We perform a $\hat U \in O(D,D)$ transformation of
the fields, and propose as an ansatz that the fields only depend on
space-time coordinates $x$ while the dependence on the internal
coordinates $\mathbb{Y}$ enters through the group element of the
transformation, $\hat U(\mathbb{Y})$. Since $\hat d$ is $O(D,D)$
invariant, it is natural to propose for it a trivial ansatz $\hat
d(x,\mathbb{Y})=d(x)$. Instead, inspired by (\ref{ODDinv}),
for the generalized metric we
propose the following decomposition:
\bea \hat {\cal H}_{\hat M\hat N}(x, \mathbb{Y})= \hat U^{\hat A}{}_{\hat
M}(\mathbb{Y})
~\hat {\cal H}_{\hat A\hat B} (x)~\hat U^{\hat B}{}_{\hat N}(\mathbb{Y})\,
,\label{Ansatz}
\eea
which in terms of the vielbein  reads
\bea \hat {\cal  E}^{\hat A}{}_{\hat M}(x,
\mathbb{Y})=
\hat {\cal
E}^{\hat A}{}_{\hat B} (x)~ \hat U^{\hat B}{}_{\hat M}(\mathbb{Y})~
\label{AnsatzVielbein}\, .\eea
In triangular gauge (\ref{TriangularGauge}),  equation
(\ref{AnsatzVielbein}) can be rewritten  as
\be \left(\begin{matrix} (\hat e^{-1})^{\hat \mu}_{~\hat m}
(x, \mathbb{Y})     & -(\hat
e^{-1})^{\hat \rho}_{~\hat m}(x, \mathbb{Y}) \hat B_{\hat \rho\hat\nu}(x,
\mathbb{Y})        \cr
                                                               0
  &
     \hat e^{\hat n}_{~\hat \nu}(x, \mathbb{Y})        \end{matrix}\right)=
\left(\begin{matrix}
(\hat e^{-1})^{\hat p}_{~\hat m}(x)     &
    -(\hat e^{-1})^{\hat r}_{~\hat m}(x) \hat B_{\hat r\hat q}(x)        \cr
                                                               0           &
     \hat e^{\hat n}_{~\hat q}(x)                    \end{matrix}\right)
                             \times \hat U(\mathbb{Y}) \, .\label{Vielbein}
\ee
Parameterizing the twist in terms of a diffeomorphism $\tilde
e(\mathbb{Y})$ and $B$-transformations $\tilde B (\mathbb{Y})$,
namely
\bea \hat U (\mathbb{Y})=\left(\begin{matrix} (\tilde e^{-1})^{\hat \mu}_{~\hat
p}(\mathbb Y) & -(\tilde
e^{-1})^{\hat \rho}_{~\hat p}(\mathbb Y) \tilde B_{\hat\rho\hat\nu}(\mathbb Y) \cr
 0      &      \tilde e^{\hat q}_{~\hat \nu}(\mathbb Y)
\end{matrix}\right)\, ,\label{hat U} \eea  we find
that the ansatz gives the following $D$-bein and Kalb-Ramond
$D$-dimensional fields \bea \hat e(x, \mathbb{Y})&=&\hat e(x)~
\tilde e(\mathbb{Y})\, ,\cr \hat B(x, \mathbb{Y})&=& \tilde
e^T(\mathbb{Y})~ \hat B(x)~ \tilde e( \mathbb{Y})+ \tilde
B(\mathbb{Y})\, . \eea

Following the organizing principle of \cite{Scherk:1979zr}, we
impose that 
 fields with only tangent space-time
indices are independent of the
internal coordinates and this implies the following block decomposition
\bea
\tilde e=\left(\begin{matrix}          1          &       0          \cr
                                      0        &      e
\end{matrix}\right) \ , \ \ \ \ \ \ \  \tilde
                                      B=\left(\begin{matrix}           0
 &       -W^T          \cr
                                      W        &       B
\end{matrix}\right)\, ,
\eea
where $e(\mathbb{Y})$ and $B(\mathbb{Y})$ are $d\times d$ matrices and
$W(\mathbb{Y})$ is a $d \times n$ matrix.
By explicitly writing the dependence on the internal coordinates  for the fields
of the effective $n$-dimensional
theory, we have
\bea
g(x, \mathbb{Y})&=&g(x) \ , \ \ \   \ \ \ \ \ \ \ \  \ \  \ \ \ \ \ \ \ \  \
\breve {B}(x, \mathbb{Y})\ = \ \breve{B}(x)\, ,
\cr
{\cal A}(x, \mathbb{Y}) &=& e^{-1}(\mathbb{Y}){\cal A}(x) \ , \ \ \   \ \ \ \  \
\ \ \ \breve{\cal B}(x, \mathbb{Y}) \
 = \ e^T(\mathbb{Y})\breve{\cal B}(x)+ W(\mathbb{Y})\, ,\cr
h(x, \mathbb{Y})&=& e^T(\mathbb{Y}) h(x)e(\mathbb{Y}) \ , \ \ \   \ \ \ \ b(x,
\mathbb{Y})\ =\
e^T(\mathbb{Y}) b(x) e + B(\mathbb{Y})\, .\eea
Note that
 $W$ involves space-time indices, so the ansatz
produces a twist in space-time explicitly breaking Lorentz invariance.
Therefore,
from now on we take $W = 0$.

Due to the length of the computation, we show how to obtain  the effective
action in two steps. In the first one,
we plug the ansatz (\ref{Ansatz}) into the action of DFT (\ref{action}).
As usual,  to have simple gauge
transformation properties,
 the reduced degrees of freedom require the following redefinitions
in terms of the original higher dimensional fields
\bea B_{\mu\nu}&=& \breve{B}_{\mu\nu}
-\frac{1}{2}\left({\cal A}^a{}_\mu {\cal B}_{a\nu}
- {\cal A}^a{}_\nu {\cal B}_{a\mu} \right)
          - {\cal A}^a{}_\mu {\cal A}^b{}_\nu b_{ab}\, ,\cr
{\cal B}_{a\mu}&=&\breve{\cal B}_{a\mu}- b_{ab}{\cal A}^b{}_\mu\,
.\label{B} \eea
 After some algebra, and using 
(\ref{strong constraint}), we obtain\footnote{Since 
the twist matrix is not necessarily globally well defined,
gauge invariance of the
original action (\ref{action}) 
requires the additional constraint $\partial_M(U^{-1})^M{}_A=0$,
which we assume from now on.}
\bea
S_{eff}&=&\int d^nx \sqrt{|g(x)|}e^{-2\Phi(x)} \left\{\vphantom{\frac12}
4g^{\mu\nu}(x)\left[\partial_\mu\partial_\nu d(x)-\partial_\mu d(x)\partial_\nu
d(x) \right]
\right.\label{FirstStep}\\
&& +~
 4~\partial_\mu g^{\mu\nu}(x)\partial_\nu d(x)-\partial_\mu\partial_\nu
g^{\mu\nu}(x)\nn\\
&&+~\frac18\left[g^{\mu\nu}(x)\partial_\mu\hat{\cal H}^{\hat A\hat B}(x) -
4\hat{\cal H}^{\hat A\mu}(x)
\partial_\mu\hat{\cal H}^{\hat B\nu}(x)\right]\partial_\nu\hat{\cal H}_{\hat
A\hat B}(x)
\cr
 && -~\frac12 F^C{}_{AB}~\hat{\cal H}^{B\hat D}(x) ~\hat{\cal H}^{A\mu}(x)
\partial_\mu\hat{\cal H}_{C\hat D}(x)\cr
 && \left.-~\frac14 F^C{}_{DA}~F^{D}{}_{CB}\hat{\cal
H}^{AB}(x)-\frac1{12}F^E{}_{AC}~F^F{}_{BD}\hat{\cal H}^{AB}(x) ~
\hat{\cal H}^{CD}(x) ~\hat{\cal H}_{EF}(x)\right\}\, .\nn
\eea
The dilaton $\Phi(x)$ is given by $e^{-2\Phi(x)}=|{\rm det}~ \hat
e(x)| ~ \tilde V
V_I~ e^{-2\hat\phi(x)}$, where $\tilde V$ stands for the volume of the dual
space-time and
we have defined the invariant
internal volume $V_I=
\int d{\mathbb Y} ~{\rm det} ~U=\int  d\mathbb{Y}$. The field $d(x)$ on the
other hand is defined by $e^{-2d(x)}= \sqrt{|g(x)|}e^{-2\Phi(x)}$.

The information on the twist
in (\ref{FirstStep}) becomes manifest only through the appearance of fluxes
$F_{ ABC}$, which are given by
\bea
F_{ABC} =  3  {\cal I}_{ D[ A} ( U^{-1})^{  M}{}_{ B} (  U^{-1})^{ N}{}_{ C]}
                                \partial_{ M}  U{}^{D}{}_N\, ,\label{Fluxes}
\eea
where $U^A_{~M}:=\hat U^A_{~M}$.
They carry $O(d,d)$ indices which are raised and lowered with the $O(d,d)$
invariant metric
\be {\cal I} = \left(\begin{matrix}0& 1_d \\ 1_d & 0\end{matrix}\right)\,
,\label{metricO66}\ee
and, being completely antisymmetric,
 they belong to the $2d  (2d-1)(2d-2) / 3!$ representation of $O(d,d)$.
Notice that, since $U\in O(d,d)$, then $( U^{-1})^{  M}{}_{ A}=U_{
A}{}^{M}$.  All the information on the twist $ U(\mathbb{Y})$ in
(\ref{hat U}) is now encrypted in the fluxes, which must be constant
in order for the effective theory to be gauge invariant. Their
presence explicitly breaks $O(d,d)$ invariance and gauges a subgroup
of it, as we shall see.

Next, we move to the second step  where we write the action 
(\ref{FirstStep}) in
terms of the physical
degrees of freedom ($i.e.$ in definite representations of Lorentz and $O(d,d)$
groups). With this aim  it appears 
 convenient to specify
 the relation between the hatted generalized metric and the physical
fields. Namely
 \bea \hat {\cal
H}^{AB}(x)&=&{\cal H}^{AB}+A^A{}_\mu g^{\mu\nu}A^B{}_{\nu},\cr \hat
{\cal H}^{A\nu}(x)&=& A^A{}_\mu g^{\mu\nu},\cr \hat {\cal
H}_{A\nu}(x)&=& -{\cal H}_{AB}A^B{}_\nu - A_A{}_\rho
g^{\rho\mu}\left(B_{\mu\nu}+\frac12 A^B_{\mu} A_{B\nu} \right),\cr
\hat {\cal H}^{\mu\nu}(x)&=& g^{\mu\nu},\cr \hat {\cal
H}^{\mu}{}_\nu(x)&=&-g^{\mu\rho}\left(B_{\rho\nu}+\frac12 A^B_{\rho}
A_{B\nu}\right),\cr \hat {\cal H}_{\mu\nu}(x)&=&g_{\mu\nu}+
A^{A}{}_\mu {\cal H}_{AB} A^B{}_{\nu} +\left(B_{\rho\mu} +\frac12
A^B_{\rho} A_{B\mu}\right) g^{\rho\sigma}
\left(B_{\sigma\nu}+\frac12 A^B_{\sigma} A_{B\nu}\right)\label{redu} 
\eea
where the  physical degrees of freedom are encoded as :
\begin{itemize}
\item {\bf Scalars} are arranged in the $O(d,d)$ covariant scalar matrix
\bea
{\cal H}_{ AB}=\left(\begin{matrix} h^{ab}
&        -h^{ac} b_{cb}      \cr
                                     b_{ac} h^{cb}
&        h_{ab}-b_{ac}h^{cd}b_{db}
                                     \end{matrix}\right)\, .
\eea
There are
 $d(d+1)/2 + d(d-1)/2$ real scalars in total,
each coming from the internal components of the $D$-dimensional
metric and $B$-field, respectively. In addition there is also the dilaton
$\Phi$.
\item {\bf Vectors} are formed by the $d + d$ vectorial degrees of freedom
coming from the metric and
$B$-field, respectively (see (\ref{VectoresDesde}) and (\ref{B})). They are
arranged in a
fundamental  representation of $O(d,d)$ as
    \bea
A^{A}{}_{\mu}=\left(\begin{matrix}{\cal B}_{a\mu}\cr -
{\cal A}^{a}{}_{\mu}\end{matrix}\right)\, .
\eea
\item {\bf Tensors} are given by the $n$-dimensional $2$-form field $B_{\mu\nu}$
and
the $n$-dimensional metric $g_{\mu\nu}$.
\end{itemize}

After some algebra, the effective action can be  finally written in a
standard form as (see the set of equations (\ref{redu}))
\bea
S_{eff}&=&\int d^nx \sqrt{|g(x)|}e^{-2\Phi(x)} \left\{ {\bf R} ~+ ~
4 ~\partial ^\mu \Phi\partial_\mu \Phi -~
\frac14  {\cal H}_{AB}{\cal F}^{A\mu\nu}
{\cal F}^{B}{}_{\mu\nu} \right. \label{our}\\
&&~~~~~~~~~~~~~~~~~~~~~~~
 -~\frac1{12}g^{\mu\nu}g^{\rho\sigma}g^{\lambda\tau}{\cal G}_{\mu\rho\lambda}
{\cal G}_{\nu\sigma\tau}
+~ \frac18 D_{\mu} {\cal H}_{AB}D^{\mu}{\cal H}^{AB}
\nn\\
 && \left. ~~~~~~~~~~~~~~~~~~~~~~~
-~~ \frac14 F^C{}_{DA}~F^{D}{}_{CB}
{\cal H}^{AB}
 - \frac1{12}F^E{}_{AC}~F^F{}_{BD} {\cal H}^{AB} ~{\cal H}^{CD} ~
{\cal H}_{EF}\right\}\, .\nn \eea Here ${\bf R}$ is the
$n$-dimensional Ricci scalar, and we have defined the field
strengths as \bea {\cal F}^{A}{}_{\mu\nu}&=&\partial_\mu
A^{A}{}_{\nu}-\partial_\nu A^{A}{}_{\mu} -F^A{}_{BC}
A^B{}_\mu~A^C{}_\nu \, ,\cr {\cal G}_{\mu\rho\lambda}&=&3
\partial_{[\mu} B_{\rho\lambda]}  -F_{ABC} A^A{}_\mu A^B{}_{\rho}
A^C{}_{\lambda}
 + 3 \partial_{[\mu}A^A{}_{\rho}A_{\lambda]A} ,
\eea 
and a covariant derivative for scalars as 
\bea D_{\mu} {\cal
H}_{AB}=\partial_\mu {\cal H}_{AB}- F^C{}_{AD}A^D{}_\mu {\cal
H}_{CB} - F^C{}_{BD}A^D{}_\mu {\cal H}_{AC}\, . \eea

The effective theory (\ref{our}) has remnant symmetries that are
inherited from the global symmetry (\ref{ODDinv}) and the gauge
invariance (\ref{gauge2D})  of its parent DFT. Concerning the
former, although the effective theory is written in a manifestly
covariant $O(d,d)$ form, the fluxes explicitly break $O(d,d)$
invariance, which is only recovered if they are treated as spurionic
fields. With respect to gauge invariance, given a $2D$-dimensional
gauge parameter $\hat \xi^{ \hat M} $, the gauge transformations
(\ref{gauge2D}) can be twisted as \bea \hat\xi^{\hat
M}(x,\mathbb{Y})=\left(\hat U^{-1}(\mathbb{Y})\right)^{\hat
M}{}_{\hat A}~ \xi^{\hat A}(x)\, , \eea and, when evaluated on the
C-bracket (\ref{Cbracket}),  the following effective algebra is
induced \bea
\left[\delta_{\xi_1},\delta_{\xi_2}\right]_C=\delta_{\xi_3} \ , \ \
\ \ \ \ \xi_3^{A}=-F^{ A}{}_{ BC}~ \xi^B_2\xi^{C}_1 \,
,\label{gauge4dim} \eea where the structure constants are given by
the fluxes (\ref{Fluxes}). In terms of the gauge parameter
$\xi^A(x)$ the fields transform as \bea \delta_\xi
g_{\mu\nu}(x)&=&0\, ,\cr
\delta_\xi d(x)& = & 0\, ,\cr
\delta_\xi B_{\mu\nu}(x)&=&-\frac12\left(A^A{}_\mu\partial_\nu \xi_A -
A^A{}_\nu\partial_\mu \xi_A\right)~,\cr
\delta_{\xi}{A}^A{}_{\mu}(x)&=&-\partial_\mu\xi^A(x) -
F^{A}{}_{BC}~\xi^B(x){A}^C{}_{\mu}~,\cr
\delta_{\xi}{\cal H}_{AB}(x)&=&-F_{AC}{}^D~\xi^C(x){\cal H}_{DB}(x)
- F_{BC}{}^D~\xi^C(x){\cal H}_{AD}(x)\, ~. \eea

The twists define the infinitesimal generators of this symmetry group, namely
\begin{equation}
{\cal E}_A = (U^{-1})_{A}^{\ \ P} {\cal E}_P\, ,
\end{equation}
which satisfy the following C-algebra
\begin{equation}
\left[{\cal E}_A, \ {\cal E}_B\right]_C =  F^D{}_{AB}\ {\cal E}_D\,
.
\end{equation}
The Jacobiator  associated to
the C-bracket does not automatically vanish, so it is not \textit{a priori}
evident whether the fluxes satisfy Jacobi
identities or not.
To answer this question we note that the C-bracket differs from the D-bracket
(which, although non skew-symmetric, does satisfy Jacobi) by a total
derivative \cite{Hohm:2010pp}
\begin{equation}
[A,\ B]_D^R =  [A,\ B]_C^R + \frac{1}{2} \partial^R (A^N B_N)\, .
\end{equation}
When particularly evaluated on the twists $ U (\mathbb{Y})$, the
last term vanishes
\begin{equation}
\partial^R \left(( U^{-1})^{M}{}_{ A} {{\cal I}}_{MN} ( U^{-1})^{N}{}_{B}\right)
= \partial^R {{\cal I}}_{AB} = 0\, ,
\end{equation}
restoring anti-symmetry and leading to a vanishing Jacobiator.
We thus conclude that the fluxes must be constrained by the usual
Jacobi identities
\begin{equation}
F^C_{\ \ [ AB} F^E_{\ \ D]C} = 0 \, .\label{JacobiFlux}
\end{equation}

\section{The effective action of DFT
 and gauged ${\cal N} = 4$ supergravity}
\label{gaugedsugra}

Setting $n = 4$ and $d = 6$, the effective action of DFT derived in
the previous section reduces to a four dimensional gauge theory with
an $O(6,6)$ global invariance. Interestingly enough, $O(6,6)$ is the
symmetry group of four dimensional gauged ${\cal N} = 4$
supergravity. Therefore, even if the starting DFT does not include
supersymmetry, we could expect a connection with the bosonic sector
of gauged supergravity. Indeed, in this Section
 we show that the effective action of DFT
 reproduces
 the electric
bosonic sector of four dimensional gauged ${\cal N} = 4$
supergravity.

We begin by reviewing this theory in the formulation of Schon and Weidner
 \cite{Schon:2006kz}
and then establish the precise correspondence. The notation used in
 \cite{Schon:2006kz} differs from ours, and the  relation
between both conventions is specified in  Table 1 and in equation (\ref{rel})
below.

\subsection{Review of gauged ${\cal N} = 4$ supergravity}

The $4$-dimensional ungauged ${\cal N} = 4$ supergravity has a global symmetry
$G = SL(2)\times
O(6,6+N)$, whose maximal
compact subgroup $K = U(1)\times O(6)\times O(6 + N)$ is realized on-shell as a
gauge symmetry of the theory.
The part of its field content that will be relevant for our discussion is
arranged in a gravity multiplet and
$6 + N$ vector multiplets involving the following fields:

\begin{itemize}
\item The {\bf gravity multiplet}
contains the metric $g_{\mu\nu}$, six
 vectors $B_{m\mu}$ ($m = 1,...,6$) and one complex scalar $\tau$
parameterizing the coset $SU(2)/U(1)$. It is useful to define the $K$ invariant
matrix
\be M_{\alpha\beta}  = \frac{1}{{\rm Im} \tau}\left(\begin{matrix}|\tau|^2 &
{\rm Re}\tau\\{\rm Re}
\tau& 1 \end{matrix}\right)\nn\ee
with $SL(2)$ indices $\alpha, \beta = \pm$. In addition, it contains four
 gravitinos
 $\psi^i_\mu$ and four spin-$1/2$ fermions $\psi^i$ ($i = 1,...,4$). The
fermions are singlets under $G$
and rotate among each other through $SU(4)\approx O(6) \in K$.

\item Each {\bf vector multiplet} contains one vector $A^a_{\mu}$ ($a$ labels
the vector multiplet), six real scalars
and four spin-$1/2$ gauginos $\lambda_a^i$. Again, the index $i = 1,...,4$ is a
gauge index rotated by $O(6) \in K$,
and the vector multiplets are rotated among each other by $O(6+N) \in K$. All
the $6 \times (6+N)$ real scalars
parameterize the coset
\be {\cal V}_M^{\ \ A} \in \frac{O(6,6+N)}{O(6)\times O(6+N)}\, ,
\ee with $M$ an $O(6,6+N)$ index and $A$ splits into $(m,a)$
with $m$ an $O(6)$ index and $a$ an $O(6 + N)$ index. This vielbein is an
element of $O(6,6+N)$ so it satisfies
\begin{equation}\eta_{MN} = - {\cal V}_M^{\ \ m}{\cal V}_N^{\ \ m} + {\cal
V}_M^{\ \ a}
{\cal V}_N^{\ \ a} \ , \ \ \ \ \ \ \eta_{MN} = {\rm diag}(------,+...+)\, .
\label{MetricSW}\end{equation}As for $\tau$, one can construct
a $K$-invariant scalar matrix
\begin{equation}
M_{MN} = {\cal V}_M^a{\cal V}_N^a + {\cal V}_M^m {\cal V}_M^m\, .
\end{equation}
\end{itemize}
Finally, the $6$ vectors of the gravity multiplet together with the $6+N$
vectors of the vector multiplets combine to form the electric vector field in
the fundamental of $O(6,6+N)$
\be
A_\mu^{\ \ M+} = \left(\begin{matrix}B_{m\mu}\\ A^a_{\mu}\end{matrix}\right)\, .
\ee

In the description of \cite{Schon:2006kz}, the ungauged theory contains the
metric $g_{\mu\nu}$, electric vector
fields $A_\mu^{\ \ M+}$ and scalars $\tau$ and $M_{MN}$ as free fields in
the Lagrangian, while the dual
magnetic vectors $A_\mu^{\ \ M-}$ and two-form gauge fields $B_{\mu\nu}^{MN}$
and $B_{\mu\nu}^{\alpha\beta}$
are only introduced on-shell. The gauging of the global group is
parameterized
by the  embedding tensors in the following representations of $G$
\begin{equation}
f_{\alpha MNP} \in (\mathbf{2},\mathbf{220}) \ , \ \ \ \ \ \ \xi_{\alpha M} \in
(\mathbf{2},\mathbf{12})\, .
\end{equation}
When these gaugings are turned on, some of the fields that were absent off-shell
in the ungauged theory are now present.

The bosonic Lagrangian is a sum
 of three terms ${\cal L}_{\rm bos} = {\cal L}_{\rm kin}+{\cal L}_{\rm
pot}+{\cal L}_{\rm top}$ given by
\bea
e^{-1} {\cal L}_{\rm kin} &=& \frac{1}{2}\ \mathbf{R} + \frac{1}{16}
D_{\mu}M_{MN} D^\mu M^{MN} + \frac{1}{8}
D_\mu M_{\alpha\beta}D^\mu M^{\alpha\beta}\cr
&&\ \ \ \ \ \ \ \ \ \ \ \ \ - \frac{1}{4} {\rm Im}(\tau) M_{MN} {\cal
H}_{\mu\nu}^{ M+}{\cal H}^{\mu\nu N +}
+ \frac{1}{8} {\rm Re} (\tau)\ \eta_{MN}\ \epsilon^{\mu\nu\rho\lambda} {\cal
H}_{\mu\nu}^{M+}
{\cal H}^{ N +}_{\rho\lambda} \, ,\cr\cr
e^{-1} {\cal L}_{\rm pot} &=& - \frac{g^2}{16}\left\{f_{+ MNP} f_{+ QRS} M^{++}
\left[\frac{1}{3}M^{MQ}M^{NR}M^{PS}
+ \left(\frac{2}{3} \eta^{MQ} - M^{MQ}\right)\eta^{NR}\eta^{PS}\right]\right.
\cr
&& \left.\ \ \ \ \ \ \ \ \ \ \ \ \ \ \ \ \ \ \ \ \ \ \ \ \ \ \ \ \ \ \ \ \ \ \ \
\ \ \ \ \ \ \ \ \ \ \ \ \ \ \ \ \ \ \
\ \ \ \ \ \ \vphantom{\frac13} + 3 \xi_{+}^M \xi_{+}^N M^{++} M_{MN}\right\}
+\dots\, ,\cr\cr
e^{-1} {\cal L}_{\rm top}&=& - \frac{g}{2}
\epsilon^{\mu\nu\rho\lambda}\left\{\xi_{+M}\eta_{NP} A_\mu^{M-} A_\nu^{N+}
A_\lambda^{P+} - \frac{g}{4}\hat f_{+ MNR} \hat f_{+ PQ}^{\ \ \ \ \ \ R}
A_\mu^{M +} A_\nu^{N+}A_\rho^{P +}
A_\lambda^{Q-}\right.\cr
&& \left.\ \ \ \ \ \  \ \ \ \ \ \ \ \ \ \ \ -\frac{1}{4} \xi_{+M}
B^{++}_{\mu\nu} \left(2 \partial_\rho
A_\lambda^{M-} - g \hat f_{+ QR}^{\ \ \ \ \ M} A^{Q+}_\rho
A^{R-}_\lambda\right)\right\} + ...\, ,
\eea
where the dots involve terms containing magnetic gaugings $f_{-MNP}$ and
$\xi_{-M}$. The covariant derivatives are
defined by
\bea
D_{\mu}M_{\alpha\beta} &=& \partial_\mu M_{\alpha \beta} + g A^{M \gamma}_\mu
\xi_{(\alpha M} M_{\beta)\gamma} -
g A^{M\delta}_\mu \xi_{+M} \epsilon_{\delta(\alpha } M_{\beta)-} + ...\, ,\cr
D_{\mu} M_{MN} &=& \partial_\mu M_{MN} + 2 g A_\mu^{P+} \Theta_{+ P(M}^{\ \ \ \
\ \ Q} M_{N)Q} +...\, ,
\eea
the field strength  by
\begin{equation}
{\cal H}_{\mu\nu}^{M+} = 2 \partial_{[\mu} A_{\nu]}^{M+} - g \hat f_{+NP}^{\ \ \
\ \ M}
A_{[\mu}^{ N\alpha} A_{\nu]}^{ P +}+ \frac{g}{2} \xi_+^M B^{++}_{\mu\nu}+ ...\,
,
\end{equation}
and also the following combinations of gaugings were defined
\bea
\Theta_{+MNP} = f_{+MNP} - \xi_{+[N} \eta_{P]M}\ , \ \ \ \ \ \ \hat f_{+ MNP} =
f_{+MNP} -
\xi_{+[M}\eta_{P]N} - \frac{3}{2} \xi_{+N}\eta_{MP}\, .
\eea

Finally, we mention that the gaugings must satisfy the following consistency
quadratic constraints

\bea
\xi^M_+ \xi_{+M} &=& 0 \, ,\\
\xi^P_+ f_{+PMN} &=& 0\, , \\
3 f_{+R[MN}f_{+PQ]}^{\ \ \ \ R} + 2 \xi_{+[M}f_{+NPQ]} &=& 0 \, , \ \ \ \ \ ...
\label{quadratic}
\eea

\subsection{Comparison with the effective action of DFT}

Having reviewed the on-shell formulation of ${\cal N}=4$
gauged supergravity, let us proceed to
compare
it with (\ref{our}) in the particular case $n=4, d=6$. First we note that
since the effective action of DFT has no $SL(2)$ symmetry mixing electric with
magnetic sectors, we can only hope to
reproduce one of these sectors, which we take to be the electric one. Since the
vectors in the effective action of DFT
arise only from the metric and B-field,
we must set $N = 0$. We then show how 
to relax
this restriction 
and finally, by gauging the rescaling symmetry, we propose
a way to obtain the gaugings $\xi_{+M}$.

Note that the $O(6,6)$ metrics employed in
(\ref{metricO66}) and
(\ref{MetricSW}) differ,
the relation between them being
\begin{equation}
R^T \eta R = {\cal I} \ , \ \ \ \ \ R =
\frac{1}{\sqrt{2}}\left(\begin{matrix}1&-1\\1&1\end{matrix}\right) \in SO(12)\,
.\label{rel}
\end{equation}
Therefore, all the comparisons made in this section are only valid up to a
rotation by this matrix. We implicitly
assume this fact.

\subsubsection{The $O(6,6)$ sector}

The fluxes in (\ref{Fluxes}) belong to the $\bf 220$ of $O(6,6)$. It
is then natural to identify them with the electric gaugings and  we
must then turn off all the magnetic fluxes $f_{-MNP}$ and
$\xi_{-M}$. There is no other source of deformations so far, so we
must also set $\xi_{+M} = 0$ (the inclusion of the full set of
electric gaugings is discussed in the forthcoming subsections).

To perform the comparison, the effective action of DFT (\ref{our}) must 
first be
taken to the
Einstein frame. The transformation $g_{\mu\nu} \to e^{2\Phi} g_{\mu\nu}$ gives

\bea
S_{eff}&=&\int d^4x \sqrt{|g(x)|} \left\{ \frac 12{\bf R}+~
\frac1{16} D_{\mu} {\cal H}_{AB}D^{\mu}
{\cal H}^{AB} -~\frac18  e^{-2\Phi}
{\cal H}_{AB}{\cal F}^{A\mu\nu}
{\cal F}^{B}{}_{\mu\nu}\right.\label{ef}\\
&& ~~~~~~~~~~~~~~~ \left. -
~\partial ^\mu \Phi\partial_\mu \Phi
 -\frac1{24}g^{\mu\nu}g^{\rho\sigma}g^{\lambda\tau}
e^{-4\Phi}{\cal G}_{\mu\rho\lambda}{\cal G}_{\nu\sigma\tau}
\right.\nn\\
 &&\left.
 ~~~~~~~~~~~~~~~
-~~e^{2\Phi} \frac18 F^C{}_{DA}~F^{D}{}_{CB}
{\cal H}^{AB}
 - e^{2\Phi}
\frac1{24}F^E{}_{AC}~F^F{}_{BD} {\cal H}^{AB} ~{\cal H}^{CD} ~
{\cal H}_{EF}\right\} \, .\nn
\eea

Then, one should identify
\begin{center}
\begin{tabular}{c|c}
Effective DFT action & Gauged ${\cal N} = 4$ sugra \\
\hline
$F_{ABC}$ & $ g f_{+MNP}$ \\
${\cal H}_{AB}$ & $M_{MN}$\\
$e^{-2\Phi}$ & $2 {\rm Im} (\tau)$\\
$A^A_{\mu}$ & $A^{M+}_{\mu}$
\end{tabular}\\
\bigskip
Table 1
\end{center}
and this guarantees that both actions and also their respective constraints
coincide.
The following subtleties must be taken into account when performing the
comparison:
\begin{enumerate}
\item Due to quadratic constraints (\ref{quadratic}), when $f_{-MNP} = \xi_{-M}
=
\xi_{+M} = 0$, the topological terms vanish ${\cal L}_{\rm top} =
0$. To see this, note that
\be f_{+R[MN}f_{+PQ]}^{\ \ \ \ \ \ R} = 0 \ \ \ \ \Leftrightarrow \
\ \ \ f_{+R[MN}f_{+P]Q}^{\ \ \ \ \ \ R} = 0\, .\ee
\item
The  term $g^2 f_{+MNP}f^{+MNP}$ in ${\cal L}_{\rm pot}$
can be explicitly written in terms of the twist, by using (\ref{Fluxes}), as
\bea
\frac 13 F_{ABC}F^{ABC}
&=&\partial_M U^A{}_N
\partial^M(U^{-1})^N{}_A\cr
&=&-\frac 12\left(\partial_M\partial^M U^A{}_N\right) (U^{-1})^N{}_A -
\frac 12
U^A{}_N \left(\partial_M\partial^M (U^{-1})^N{}_A \right) \eea and
therefore it vanishes due to the level matching condition. This in turn
implies that the effective action (\ref{ef}) is in fact a truncation
of an ${\cal N} = 8$ theory to ${\cal N} = 4$
\cite{Aldazabal:2011yz, Dibitetto:2011eu}, so only a subset of
gauged ${\cal N} = 4$ supergravities can be reached through
Scherk-Schwarz compactification from DFT.
\item To make the equivalence evident, the two-form $B_{\mu\nu}$
must be dualized in the effective action of DFT to give an
equivalent theory in terms of a  scalar field $\lambda \equiv 2{\rm
Re} (\tau)$. Introducing $\lambda$ as a Lagrange multiplier to
enforce $\epsilon^{\sigma\mu\nu\rho}\partial_\sigma \partial_\mu
B_{\nu\rho}=0$, one defines the following action for $B_{\mu\nu}$
\bea S_B = -\int \frac{d^4x}{24}{\sqrt {g}} \left [e^{-4\Phi} {\cal
G}_{\mu\nu\rho} {\cal G}^{\mu\nu\rho} +2\lambda
\epsilon^{\sigma\mu\nu\rho} \partial_\sigma ( {\cal
G}_{\mu\nu\rho}+F_{ABC}A^A_\mu A^B_\nu A^C_\rho - 3
\partial_{\mu} A^A_{\nu}A_{A\rho})\right]\, .
\nn
\eea
Here ${\cal G}_{\mu\nu\rho}$ must be eliminated through its equation of motion
\be
{\cal G}^{\mu\nu\rho}=e^{4\Phi}\epsilon^{\sigma\mu\nu\rho}
\partial_\sigma\lambda
\, ,
\label{dualaxion}
\ee
which replaced in $S_B$ gives
\bea
S_B=-\int \frac{d^4x}{24}\sqrt{g}e^{4\Phi}\left [6\partial_\sigma\lambda
\partial^\sigma\lambda
-2\lambda \epsilon^{\sigma\mu\nu\rho}\partial_\sigma \left (-
F_{ABC} A_\mu^A A_\nu^B A_\rho^C + 3\partial_{[\mu} A^A_{\nu}A_{\rho
]A}\right )\right ]\, .\nn \eea The first term contributes to the
kinetic term for the gravity scalars, and the second one coincides
with the last term in ${\cal L}_{\rm kin}$ upon use of quadratic
constraints.
\end{enumerate}

\subsubsection{Including vectors}\label{vectors}

We showed above
that the effective action of DFT  matches the bosonic electric 
sector of gauged
${\cal N}=4$ supergravity
 when the number of vector multiplets is six,
 $i.e.~ {N}=0$,
and $\xi_{+ M}=0$. Here we show how to relax the first restriction.

To increase  the number of electric vector fields, we are inspired
by the extension of DFT to the abelian heterotic string performed in 
\cite{Hohm:2011ex},
where the $2D$ coordinates $X^{\hat\mu} , \tilde X_{\hat\mu}, \hat\mu=
0,\dots , D-1$
are extended by $N$ extra coordinates $z^i$,
$i=1, \dots , N$ and,
correspondingly, the generalized metric $\hat {\cal H}_{\hat {M}\hat
{N}}$
is enlarged to a $(2D+N)\times (2D+N)$ matrix that naturally
incorporates $N$ additional vector fields $\hat V_{\hat \mu}^i$.

We will use the following  $O(D, D+N)$ metric
\bea
\hat{\cal I}_{\hat {M}\hat {N}}=       \left(\begin{matrix}
{\cal I}^{\hat\mu\hat\nu} & {{\cal I}^{\hat\mu}}_{\hat\nu} &
 {{\cal I}^{\hat \mu}}_i\cr
{{\cal I}_{\hat\mu}}^{\hat\nu} & {\cal I}_{\hat\mu\hat\nu} &
{\cal I}_{\hat \mu i}\cr
{{\cal I}_i}^{\hat\nu} & {\cal I}_{i\hat\nu} & {\cal I}_{ij}
\end{matrix}
\right)
=  \left(\begin{matrix}
     0         &  1_{D} &       0        \cr
1_{D}  &        0       &       0        \cr
     0         &        0       &   1_{N}
\end{matrix}
\right)\,.
\label{etaa}
\eea

We formally keep the action and the form of the gauge transformations
but with respect to the enlarged generalized metric
$\hat {\cal H}_{\hat {M}\hat {N}}$ introduced in \cite{Hohm:2011ex},
namely

\bea
\hat{\cal H}_{\hat M\hat N}&=&
                \left(\begin{matrix}
{\hat g}^{\hat \mu\hat \nu}
             &    -{\hat g}^{\hat \mu\hat \rho}{\hat c}_{\hat \rho
\hat \nu}   &   -{\hat g}^{\hat \mu\hat \rho}\hat V_{j\hat \rho }
\cr
 - {\hat c}_{\hat \rho\hat \mu}{\hat g}^{\hat
\rho\hat \nu}
&   {\hat g}_{\hat \mu\hat \nu}+{\hat c_{\hat \rho\hat \mu}
{\hat g}^{\hat \rho\hat \sigma}{\hat  c}_{\hat \sigma\hat \nu}}+
\hat V_{k\hat\mu}\hat V_{k\hat\nu }
&   {\hat V}_{j\hat \mu }+{\hat c_{\hat \rho\hat\mu}
{\hat g}^{\hat \rho\hat \sigma}{\hat  V}_{j\hat \sigma }}
\cr
-{\hat V}_{i\hat \rho }{\hat g}^{\hat
\rho\hat \nu}
&   {\hat V}_{i \hat \nu }+{\hat c_{\hat \rho\hat \nu}
{\hat g}^{\hat \rho\hat \sigma}{\hat  V}_{i \hat \sigma }}
&   \delta_{ij}+{\hat V_{i\hat \rho  }
{\hat g}^{\hat \rho\hat \sigma}{\hat  V}_{j\hat \sigma }}
\end{matrix}
\right)\, ,\eea
where
the indices of the  gauge group  $i, j, \dots $ are raised and lowered with
$\delta_{ij}$ and
\be
\hat c_{\hat\mu\hat\nu}=\hat B_{\hat\mu\hat\nu}+\frac 12
{\hat V_{\hat\mu}}^{~i} \hat V_{i\hat\nu }\, .
\ee
The generalized vielbein producing $\hat{\cal H}_{\hat {M}\hat {N}}
={{\cal E}^{\hat {A}}}_{\hat {M}}\hat\eta_{\hat {A}\hat {B}}
{\hat{\cal E}^{\hat {B}}}_{~~\hat {N}}$ is
\bea
\hat{\cal E}^{\hat {B}}_{~\hat {N}}=
\left(\begin{matrix}
(\hat e^{-1})^{\hat \nu}{}_{\hat n}
& -(\hat e^{-1})^{\hat\sigma}{}_{\hat n}
\hat c^{}_{\hat\sigma\hat\nu} & -(\hat e^{-1})^{\hat\sigma}{}_{\hat n}
\hat V^{}_{\hat\sigma j}\cr
0 & \hat e^{\hat n}{}_{\hat\nu} & 0\cr
0 & \hat V_{i\hat\nu } & \delta_{ij}
\end{matrix}
\right )
\, .
\eea

The compactification procedure closely follows
\cite{Kaloper:1999yr}. The metric and $B$ fields in terms of the
compactified degrees of freedom take the form (\ref{VectoresDesde}),
where now 
\bea 
B_{\mu\nu}&=&\breve B_{\mu\nu}-\frac 12\left ({\cal
A}^a_\mu B_{a\nu}- {\cal A}^a_\nu B_{a\mu}\right ) - b_{ab}{\cal
A}^a_\mu {\cal A}^b_\nu- \frac 12 \hat V^i_c \left ( {\cal A}^c_\mu
V^i_\nu -{\cal A}^c_\nu V^i_\mu\right )\, ,\cr {\cal B}_{a\mu}&=&\breve {\cal
B}_{a \mu} - {b}_{ac} {\cal A}^c_\mu -\frac 12 V^i_a V ^i_\mu \, ,
\eea
and the vector fields are 
\be 
V^i_{\mu}= \hat V_{\mu}^i - \hat V^i_c
{\cal A}^c_\mu\ . 
\ee

Generalizing the procedure of the previous section,
we define the reduced gauge multiplet as
\bea
{A}^{A}_\mu = \left (\begin{matrix}
{\cal B}_{a\mu} &\cr
-{\cal A}^a_\mu & \cr
-V^i_\mu \end{matrix}\right ) \, ,\label{rgmv}
\eea
and the reduced generalized metric as
\bea
{\cal H}_{AB} =
\left (\begin{matrix} h^{ab} & -h^{ad}c_{db}
& -h^{ad}V^j_\gamma \cr
c_{ad}h^{db} & h_{ab}-c_{ad}
h^{de}c_{eb}+ V^i_a V^i_b & V_a^j
+c_{da}h^{de}V^j_{e}\cr
- V_d^ih^{bd}
 & V^i_b+c_{db}h^{de}V_\delta^i& \delta_{ij} +
V_d^i h^{de}V^j_e
\end{matrix} \right )\, ,
\eea
where $c_{ab}=b_{ab}+\frac 12 V_a^i V_b^i$.

The global symmetry
to be twisted is now
$O(D,D+N)$ and we extend the  ansatz (\ref{Ansatz}) for the generalized
metric as
\be
\hat{\cal H}_{\hat{M}\hat{N}}(x,{\mathbb Z})=
{\hat U^{\hat{A}}{}_{\hat{M}}}({\mathbb Z})~
\hat{\cal H}_{\hat{A}\hat{B}}(x)~
{\hat U^{\hat{B}}{}_{\hat{N}}}({\mathbb Z})\, ,
\ee
where ${\mathbb Z}=({\mathbb Y}, z)$ and $\hat U\in O(D,D+N)$.
In terms of the vielbein, it
reads
\bea
\hat{\cal E}^{\hat{A}}_{~~\hat{M}}
(x,{\mathbb Z})= \hat{\cal E}^{\hat{A}}{}_{\hat{B}}(x)~
\hat U^{\hat{B}}{}_{\hat{M}}({\mathbb Z})~
\, ,
\eea
which can be rewritten in matricial form as
\bea
&&
\left(\begin{matrix}
(\hat e^{-1})^T(x,{\mathbb Z})
& -(\hat e^{-1})^T(x,{\mathbb Z}) \hat c(x,{\mathbb Z)}
& -(\hat e^{-1})^T(x,{\mathbb Z})
\hat V^T(x,{\mathbb Z})\cr
0 & \hat e(x,{\mathbb Z}) & 0\cr
0 & \hat V(x,{\mathbb Z}) & 1
\end{matrix}
\right )\cr
&&\qquad\qquad =
\left(\begin{matrix}
(\hat e^{-1})^T(x)
& -(\hat e^{-1})^T(x) \hat c(x)
& -(\hat e^{-1})^T(x)
\hat V^T(x)\cr
0 & \hat e(x) & 0\cr
0 & \hat V(x) & 1
\end{matrix}
\right )
\times\hat U({\mathbb Z})
\, .
\eea
Parameterizing the twist in terms of diffeomorphisms $\tilde e$,
$\tilde B$ and $\tilde V$ transformations,
namely
\bea
\hat U({\mathbb Z})=
\left(\begin{matrix}
 (\tilde e^{-1})^T
& - ~(\tilde e^{-1})^T~\tilde c
&  -~(\tilde e^{-1})^T \tilde V^T
\cr
0  & \tilde e &  0\cr
0
& \tilde V
& 1
\end{matrix}
\right )\, ,
\eea
the ansatz gives the following $D$-dimensional gauge fields
\bea
\hat V(x,{\mathbb Z})&=&\tilde V({\mathbb Z})+\tilde e({\mathbb Z})\hat V(x)
\, .
\eea

Now everything goes on exactly as before. The
reduced effective action has the same form of the previous section, namely
(\ref{our}),
 with the obvious extension in the values taken by
the indices and the inclusion of $N$ extra vector fields $V^i_{\mu}$ in the
definition (\ref{rgmv}).

\subsubsection{Rescaling  symmetry gaugings}

In this section we propose a way to obtain the gaugings $\xi_{+ M}$
in the $({\bf 2,12})$ representations of $SL(2)\times O(6,6)$. For
simplicity here we restrict to the $N = 0$ case, but the results can
be extended if heterotic vector multiplets are introduced as in the
previous subsection and their rescaling taken into account. To
facilitate the presentation, in this subsection we turn off the
gaugings in the $({\bf 2,220})$.

In \cite{Derendinger:2007xp}, a  rescaling symmetry of the NS-NS
sector of the heterotic theory was twisted, and it was shown that
this gives rise to the gaugings $\xi_{+m}$ with $m= 1,\dots\,6$.
This symmetry acts by rescaling  the metric and $B$-field, and by shifting
the dilaton, and it is explicitly broken in DFT due to the
dependence on the dual coordinates. However, at a formal level DFT
is invariant under a global symmetry given by
\begin{equation}
\hat d \to \hat d \pm \gamma \ , \ \ \ \ \ \hat {\cal H}^{\hat M\hat
N} \to  \hat O^{\hat M}{}_{\hat P}  \hat {\cal H}^{\hat
P\hat Q} \hat O^{\hat N}{}_{\hat Q} \ , \ \ \ \  \ \hat O  =
\left(\begin{matrix} e^\gamma & 0 \\
0 & e^{-\gamma}
\end{matrix}\right) \, ,\label{covariantrescaling}
\end{equation}
whenever the following relation holds
\begin{equation}
\hat O^{\hat P}{}_{\hat M} \partial_{\hat P} \to e^{\pm \gamma}
\partial_{\hat M}\, ,
\end{equation}
i.e., when either \be\partial^m ~\bullet = 0 \ \ \ \ \ \ \ \ (-)
\label{xiviejo} \ee or \be
\partial_m ~\bullet = 0  \ \ \ \ \ \ \ \ (+)\, . \label{xinuevo}
 \ee
While the former (\ref{xiviejo}) corresponds to the case analyzed in
\cite{Derendinger:2007xp}, the later is only possible due to the
doubling of the coordinates, and symmetry arguments suggest that
they should give rise to the $\xi_+^m$ gaugings. Here we use the
covariant notation (\ref{covariantrescaling}), but one should keep
in mind that the extension of the symmetry is only possible if
covariance is broken through either (\ref{xiviejo}) or
(\ref{xinuevo}).

The compactification procedure is equivalent to the one before. One
transforms the fields and makes the transformations carry the
 internal dependence
\begin{equation}
\hat d (x, \mathbb{Y}) = \hat d (x)  \pm \gamma (\mathbb{Y}) \ , \ \
\ \ \ \hat {\cal H}^{\hat M\hat N}(x, \mathbb{Y}) \to \hat O^{\hat
M}{}_{\hat P}(\mathbb{Y}) \hat {\cal H}^{\hat P\hat Q} (x)\hat
O^{\hat N}{}_{\hat Q}(\mathbb{Y}) \, .\label{ansatzrescaling}
\end{equation}
The $\pm$ depends on whether (\ref{xiviejo}) or (\ref{xinuevo}) is
chosen.

The comparison of the effective action with gauged ${\cal N} = 4$
supergravity is not straightforward, and one must use the
dualizations of \cite{Derendinger:2007xp}. Instead of dualizing the
$B$-field into the axion of \cite{Schon:2006kz}, one must integrate
the $A^{M-}_{\mu}$ fields out and use the local gauge invariance
under (axionic) rescaling shifts to gauge away the axion ${\rm Im}
\tau = 0$. This yields the following equivalent formulation of
gauged ${\cal N} = 4$ supergravity when only $\xi_+$ gaugings are
turned on
\begin{eqnarray}
e^{-1} {\cal L} &=& \frac{1}{2} R + \frac{1}{16} D_{\mu}M_{MN}
D^{\mu}M^{MN} - D_\mu \phi D^\mu \phi \\ &&  -\frac{1}{4} e^{-2\phi}
M_{MN} {\cal H}^{M+}_{\mu \nu}{\cal H}^{N\mu\nu +} - \frac{3}{8}
e^{-4\phi} g^2 \Omega_{\mu \nu \rho} \Omega^{\mu \nu \rho} + e^{-1}
{\cal L}_{pot}\, ,\nn
\end{eqnarray}
where
\begin{eqnarray}
D_\mu \phi &=& \partial_\mu \phi - \frac{1}{2} \xi_{+M}A^{M+}_\mu \, ,\\
D_\mu M_{MN} &=& \partial_{\mu } M_{MN} - g A_{\mu}^{P+}
\xi_{+M}M_{NP} + g A^{+}_{\mu M} M_{NP} \xi^{+P}\, ,\\
\Omega_{\mu\nu\rho} &=& \partial_{[\mu} B_{\nu\rho]}^{++} -
\xi_{+M}A^{M+}_{[\mu} B^{++}_{\nu\rho]}  - 2 A^{N+}_{[\mu}
\partial_\nu A_{\rho]N}^+\, ,\\
{\cal H}^{M+}_{\mu\nu} &=& 2 \partial_{[\mu}A_{\nu]}^{M+} - \frac g
2 \xi_{+N}A^{N+}_{[\mu} A^{M+}_{\nu]} + \frac g 2 \xi^M_+
B^{++}_{\mu\nu}\, .
\end{eqnarray}

The identifications are those of the previous section plus
 $B^{++}_{\mu\nu} \leftrightarrow B_{\mu\nu}$ and the
gaugings are originated from the following derivatives
\be \xi_{+M} \ \ \leftrightarrow \ \  \partial_M \gamma
(\mathbb{Y})\, .\ee

Equations (\ref{xiviejo})-(\ref{xinuevo}) imply that the constraints
$\xi_{+}^M\xi_{+M} = 0$ are automatically satisfied.

\section{(Non)geometric string backgrounds}\label{NonGeomSec}

In this Section  we show that the generalized Scherk-Schwarz
mechanism can be interpreted as a reduction on a twisted double
torus. The string backgrounds associated to different twists are
analyzed and their connection to standard (dual) fluxes is discussed.

\subsection{Reduction on a twisted double torus}

The dependence of the generalized metric and its vielbein on the internal
coordinates
$\mathbb{Y}^M$ was introduced in (\ref{Ansatz}) as
\begin{equation}
{\cal H}_{MN} (x,\mathbb{Y}) = U^A{}_M(\mathbb{Y})\ {\cal H}_{AB} (x)\
U^B{}_N(\mathbb{Y})\, .
\end{equation}
In the limit in which the scalar matrix approaches the identity ${\cal
H}_{AB}(x) \to \delta_{AB}$,
the generalized metric takes the form
\begin{equation}
{\cal H}_{MN} (x,\mathbb{Y}) \to H_{MN}(\mathbb{Y}) = U^A{}_M(\mathbb{Y})\
\delta_{AB}\ U^B{}_N(\mathbb{Y})\, .
\end{equation}
Sketched in this form, it is natural to identify the twist matrix $U(\mathbb{Y})$
with a vielbein for the metric
$H(\mathbb{Y})$.
In fact, this metric is invariant under $O(d)\times O(d)$ transformations that
preserve the identity and act on
the twist from the left, and  it transforms under $O(d,d)$ acting on the twists
from the right.
Therefore, the twists $U(\mathbb{Y})$ can be interpreted as generalized internal
vielbeins, associated to the
internal geometry on which we are compactifying.
Any non-trivial scalar matrix ${\cal H}_{AB}(x)$ can later be interpreted as
scalar fluctuations deforming
the background described by $H(\mathbb{Y})$.

It is useful to introduce the $1$-forms
\begin{equation}
\Gamma^A = U^{A}_{\ \ M} (\mathbb{Y})d\mathbb{Y}^M \,
.\label{oneformsGammas}
\end{equation}
When the double space is untwisted,  these  $1$-forms are simply the
differentials $d\mathbb{Y}^A$, which are globally well defined and
covariantly constant with respect to the standard derivative $d$.
The Scherk-Schwarz twisted torus \cite{Scherk:1979zr}
 has instead globally well defined
1-forms, $\Gamma^a=U^a{}_\alpha dy^{\alpha}$, which are covariantly 
constant with respect to a derivative 
that includes a constant non-vanishing Levi-Civita connection 
$\omega^a{}_b$. This becomes manifest
through the Maurer-Cartan equation
\begin{equation}
d \Gamma^a + f^a_{\ \ bc} \Gamma^b \wedge \Gamma^c = 0\, .
\end{equation}
Interestingly, $f^a{}_{bc}:=-\omega^a{}_{[bc]}$ can be defined through
the Lie bracket that in turn defines the gauge symmetry of the
effective action, and in fact, the connection is precisely given by
the fluxes. The Scherk-Schwarz twisted torus can  be conceived as
the manifold of a group, the components of $f$ being its structure
constants. Such group is the gauge group of the effective action,
where the parameters $f$ appear as gaugings.

We have seen already that many of these features are preserved in
our scenario. For instance, the standard Lie bracket is replaced by
the C-bracket which determines the gauge invariance of the effective
action. However, it is not a priori clear what is the relation
between the fluxes $F_{ABC}$ and the connection and torsion of the
double twisted torus.

The proper formalism to deal with in the context of double field theory 
was developed in \cite{Hohm:2010xe}, generalizing the frame-like 
geometrical formalism introduced by Siegel \cite{Siegel:1993}.
 It is useful to recall the following
definition of the 2-form torsion
\bea 
T^a{}(V,W)=\nabla_V W^a- \nabla_W V^a - [V,W]^a \, ,\label{torsion}
\eea 
where
$V,W$ are two vector fields, $\nabla$ denotes the covariant
derivative and $[\,  , ]$ the Lie bracket. This can alternatively be
written in the more convenient form 
\bea 
T^a{}(V,W)=[{\cal
L}_V^{\nabla}~W-{\cal L}_{V}W]^a\, , 
\eea 
where ${\cal L}_V$ is the
Lie derivative of vector fields and the upper index ${\nabla}$ means
one must change the partial derivative by the covariant one.

So, we define the generalized torsion in the context of double field theory 
as in (\ref{torsion}), where the label $a$ is replaced by $A$ and the standard 
Lie bracket is replaced by the C-bracket \cite{Hohm:2010xe} (see 
\cite{Coimbra:2011nw} for a discussion of torsion in 
the context of Generalized Geometry). Using the Lie derivative (\ref{c-LD}) and
the covariant derivative 
\bea 
{\nabla}_{M} V^A= \partial_M V^A+
\omega^A{}_{BM} V^B \, , 
\eea 
where $\omega^A{}_{B}$ represents the
generalized spin connection, and imposing the metricity condition
(which implies $\omega_{AB}=-\omega_{BA}$) one gets
\bea  T_{ABC}=\omega_{ACB}+\omega_{BAC} + \omega_{CBA} - \left({\cal
L}_{{\cal E}_B} {\cal E}_C\right)_A = -3~\omega_{[ABC]}+ F_{ABC}\, .
\eea

Different connections define different torsions. In the torsionless
case, the connection and the fluxes are related through
\begin{equation}F_{ABC} = 3~\omega_{[ABC]}  \, .\end{equation}
Notice that the antisymmetry 
of $F_{ABC}$ in all the
indices  implies
\bea
F^A{}_{AB}=0,
\eea
which generalizes the equation $f^\alpha{}_{\alpha\beta}=0$, ensuring the
invariance of the measure under the killing isometries of geometric
backgrounds with  $f$ fluxes.

\subsection{Generalized fluxes and superstring
compactifications}

The low-energy effective theories arising in standard
compactifications of string theory on manifolds with
reduced structure or
orbifolds are gauged supergravities in which only a subset of
gaugings can be obtained. Those gaugings whose origin from a
$10$-dimensional supergravity has not been identified are dubbed
non-geometric. From a  stringy perspective, the main motivation for
introducing these fluxes is duality. Interestingly, being invariant
under T-duality transformations, DFT provides a higher dimensional
framework in which all fluxes related to the antisymmetric field
strength $H$-flux through T-duality appear on an equal footing.

%%%%%%%%%%%%%%%%%%%%5XXXX
We have shown in Section 4 that the  compactified theory can be
identified with the bosonic electric sector of ${\cal N}=4$ gauged
supergravity with broken global symmetry group $G= SL(2,\mathbb{R})\times
O(6,6+N)$. From a stringy perspective, the ${\cal N}=4$ supergravity
theory could arise, for instance, from the compactification of
$D=10$ heterotic string \cite{Kaloper:1999yr}, but also from
compactifications of Type II  
either on toroidal orientifolds
\cite{Aldazabal:2008zza},\cite{Dall'Agata:2009gv} or
 on $SU(2)$-structure manifolds
\cite{Danckaert:2011ju}. The global group $G$ has different
interpretations in each case (see \cite{Aldazabal:2010ef} for
details).

In the heterotic case,  $SL(2,\mathbb{R})$ corresponds to fractional
linear transformations of the heterotic axio-dilaton 
 field  $\tau_H=
\lambda +i e^{-2\phi}$
(where $\lambda$ is the dualized 2-form
Kalb-Ramond field) whereas $O(6,6)$ (let us set $N=0$ for the
moment) is the group that contains T-duality transformations, as
introduced in the context of Generalized Geometry.

However, if we consider Type IIB compactifications,
$SL(2,\mathbb{R})$ is the S-duality group that, in particular, acts
on the IIB axio-dilaton field $\tau_B= C_0+i e^{-2\phi}$ (where
$C_0$ is the RR $0$-form) as modular transformations. T-duality
transformations are more involved here and imply the action  of
elements of both subgroups in $G_B=SL(2,\mathbb R)\times O(6,6)_B$.

The completion of T-duality transformations to the full $O(6,6)$ group was
part of the original motivation for writing down the DFT action
(\ref{action}) and, therefore, the identification with a heterotic
compactification is naturally suggested.  In particular, if  the 2-form
field ${\hat B}_{\hat \rho\hat \nu}$ in
(\ref{MetricaGeneralizada}) is the Kalb-Ramond field of the heterotic
string, we see from (\ref{dualaxion}) that $\tau = \lambda +i
e^{-2\phi}$ would indeed be the heterotic axio-dilaton field.  In
what follows we will pursue this interpretation, but we will comment
at the end of this section that a Type IIB orientifold interpretation is
seemingly possible.

We now address the issue of the $2d$-dimensional origin of the
fluxes. Starting from their definition (\ref{Fluxes}):
\begin{equation}
F_{ABC} = 3\ {\cal I}_{D[A}\ (U^{-1})^{N}{}_{B} (U^{-1})^{P}{}_{C]}
\partial_N U^D{}_P \, ,\label{FluxesAppendix}
\end{equation}
we parameterize the internal $2d$-bein in the most general form (see
Appendix \ref{beta})
\begin{equation}
U = \left(\begin{matrix} (e^{-1})^T(1 + B\beta )  &   -(e^{-1})^T B
\\ -e\beta  & e\end{matrix}\right) \ , \ \ \ \ U^{-1} = \left(\begin{matrix} e^T
&
 B e^{-1}
\\ \beta e^T & (1 + \beta B)e^{-1} \end{matrix}\right)\, .
\end{equation}
One can then define the relation between the tensor $F_{ABC}$ and
the standard set of NS-NS string fluxes of \cite{Shelton:2005cf}.
Defining for simplicity $\tilde a \equiv a + d$, we set
\begin{equation}
F_{\tilde a \tilde b \tilde c} = H_{abc} \ , \ \ \ \ F^{\tilde a}_{\
\ \tilde b \tilde c} = \omega^{ a}_{ bc}\ , \ \ \ \ F^{\tilde a
\tilde b}_{\ \ \ \ \tilde c} = Q^{ab}_c \ , \ \ \ \ F^{\tilde a
\tilde b \tilde c} = R^{abc} \, .\label{FluxDefhet}
\end{equation}

The different fluxes take the following form in terms of this
parameterization
\bea
H_{abc} &=&  3 \left\{\partial^\alpha [(1 + \beta
B)e^{-1}]^\gamma_{\ [a} (e^{-T }B)_{b\underline\alpha} (e^{-T}B)_{c]\gamma} -
\partial^\alpha(B
e^{-1})_{\gamma[a} (e^{-T}B)_{b\underline \alpha} [e^{-T} (1 + B \beta)]_{c]}^{\
\ \gamma}
\right. \nn\\ && - \partial_\alpha[(1  + \beta B)e^{-1}]^\gamma_{\  [a}
[e^{-T}(1 + B \beta)]_b^{\ \alpha}
(e^{-T}B)_{c]\gamma} \nn\\
&& \left.+ \partial_\alpha (B e^{-1})_{\gamma[a}[e^{-T }(1 + B
\beta)]_b ^{\ \alpha} [e^{-T}(1 + B \beta)]_{c]}^{\ \
\gamma}\right\}\, , \label{Hflux} \eea \bea \omega^a_{ bc} &=&  -
  \partial^\alpha (\beta e^T)^{\gamma a} [e^{-T} (1 + B
\beta)]_{[b}^{\ \ \alpha} (e^{-T} B)_{c]\gamma} -
\partial^\alpha(e^T)_\gamma^{\ a} (e^{-T} B)
_{[b\underline\alpha} [e^{-T}(1 + B \beta)]_{c]}^{\ \ \gamma}\nn\\
&& - \ \partial_\alpha (\beta e^T)^{\gamma a} [e^{-T}(1 + B \beta)]_{[b}^{\ \
\alpha} (e^{-T}B)_{c]\gamma}
+ \partial_\alpha (e^T)_\gamma^{\ a} [e^{-T}(1 + B \beta)]_{[b}^{\ \ \alpha}
[e^{-T}(1 + B \beta)]_{c]}^{\ \ \gamma} \nn\\
&& +\  e^a_{\ \alpha}\left\{  \partial^{[\alpha}[(1 + \beta B)
e^{-1}]^{\gamma]}_{\ \ [b} (e^{-T}B)_{c]\gamma}\right. \nn\\ &&\ \ \
\ \ \ \ \ \  \left. +\
\partial_\gamma[(1 + \beta B)e^{-1}]^\alpha_{\ [b}[e^{-T}(1 + B
\beta)]_{c]}^{\ \ \gamma} -  \partial^\alpha
(Be^{-1})_{\gamma[b}[e^{-T}(1 + B \beta)_{c]}^{\ \ \gamma}]\right\}
\nn\\&& -\ (e\beta)^{a\alpha}\left\{\partial_{[\gamma} (B
e^{-1})_{\alpha][b} [e^{-T}(1 + B
\beta)]_{c]}^{\ \ \gamma}\right.\nn\\
&& \ \ \ \ \ \ \ \ \ \ \ \  \left. +\ \partial_\alpha [(1 + \beta
B)e^{-1}]^\gamma_{\ \ [b} (e^{-T}B)_{c]\gamma} -   \partial^\gamma
(B e^{-1})_{\alpha[b} (e^{-T}B)_{c]\gamma}\right\}\, ,
\label{omegaflux}\eea
\bea Q^{bc}_{  a} &=&    \partial^\alpha [(1 + \beta B)
e^{-1}]^\gamma_{\ a} e^{[b}_{\ \ \alpha } e^{c]}_{\ \ \gamma} -
\partial^\alpha (B e^{-1})_{\gamma a} e^{[b}_{\ \ \alpha}
(e\beta)^{c]\gamma}\label{QFlux}\\
&& -\ \partial_\alpha [(1 + \beta B)e^{-1}]^\gamma_{\ a} (e \beta)^{[b\underline
\alpha}e^{c]}_{\ \ \gamma} + \partial_\alpha
 (Be^{-1})_{\gamma a}(e\beta)^{[b\underline \alpha}(e\beta)^{c]\gamma}\nn\\
&& - (e^{-T}B)_{a\alpha}\left\{ \partial^{[\alpha}(\beta
e^T)^{\gamma][c} e^{b]}_{\ \ \gamma}+\ \partial_\gamma (\beta
e^T)^{\alpha[c}(e\beta)^{b]\gamma} -
\partial^\alpha (e^T)_\gamma^{\ [c} (e\beta)^{b]\gamma}\right\}
\nn\\
&& + [e^{-T}(1 + B \beta)]_a^{\ \alpha} \left\{\partial_{[\gamma}(e^T)_{\alpha
]}^{\
\ [c} (e\beta)^{b]\gamma}+\ \partial_\alpha (\beta e^T)^{\gamma[c}e^{b]}_{\ \
\gamma} -
  \partial ^\gamma (e^T)_\alpha^{\ [c} e^{b]}_{\ \ \gamma}\right\}
\, , \nn\\&&\nn\\
R^{abc} &=&  3e^a{}_\alpha e^b{}_{\lambda} e^c{}_\gamma 
\left\{ -\partial_\delta \beta^{[\alpha\lambda}\beta^{\gamma]\delta} 
+ \partial^{[\alpha}\beta^{\lambda\gamma]}\right\},\label{RFlux} 
\eea
where underlined indices must not be antisymmetrized.

 It is interesting to analyze the dependence of the  fluxes in different limits
of interest:
\begin{itemize}
\item 
Setting $\beta = 0$ and
taking $B$, $e$ 
to depend only on base coordinates, the $H$-flux takes the geometric form
\begin{equation}
H_{abc} = 3 (e^{-1})^\alpha{}_a(e^{-1})^\beta{}_b(e^{-1})^\gamma{}_c
\partial_{[\alpha} B_{\beta\gamma]} \, .\label{Hsimple}
\end{equation}
Gauge transformations for $B \to B + d\Lambda$ are given
by $O(d,d)$ $B$-type transformations acting from the right
\bea
\left(\begin{matrix}( e^{-1})^T & - (e^{-1})^T B\\ 0 & e
\end{matrix}\right)  &\to &
   \left(\begin{matrix}(e^{-1})^T & - (e^{-1})^T B \\ 0 & e
\end{matrix}\right) \left(\begin{matrix}1 & -d \Lambda \\ 0 & 1
\end{matrix}\right)  \cr
&&=  \left(\begin{matrix}(e^{-1})^T & -e^{-1}(B + d \Lambda)   \\ 0 & e
\end{matrix}\right)\, , \nn\eea
and it can be seen from (\ref{Hsimple}) that  they leave the
gaugings invariant.

\item 
Setting $B= \beta = 0$ and taking $e$ to
depend only on base coordinates, the $\omega$-flux takes the simple form
\begin{equation}
\omega^a_{ bc} =  2  (e^{-1})^\alpha_{\ [b}(e^{-1})^\gamma_{\ c]}
\partial_\alpha
e^a_{\ \gamma}\, . \label{omegaSS}
\end{equation}
Or also, by letting $B$ depend on dual coordinates and setting $e =
{\bf 1}$ and $\beta = 0$, one gets
\begin{equation}
\omega^a_{bc} = \partial^a B_{bc} \, .\label{omegaB}
\end{equation}
\item 
Setting $e = {\bf 1}$, $B= 0$ and $\beta$
depending only on base coordinates, the $Q$-flux reads
\begin{equation}
Q^{ab}_{ c} =  \partial_c \beta^{ab}\, . \label{Qbeta}
\end{equation}
Also  setting $\beta = B = 0$ and making $e$ depend on dual
coordinates only, one gets
\begin{equation}
Q^{ab}_c = 2 e^{[a}_{\ \ \alpha} e^{b]}_{\ \ \gamma}
\partial^\alpha (e^{-1})^\gamma_{\ c}\, . \label{QSS}
\end{equation}
Notice that, from the first and third terms in (\ref{QFlux}) and
keeping only ordinary derivatives, we obtain $Q^{ab}_{ c}=\beta ^{r
a}\beta^{bs}H_{rsc}+..$ reproducing the expression found in 
\cite{Grana:2008yw} (see also \cite{Andriot:2011uh}).

\item The $R$-flux can be obtained by setting  $e = {\bf 1}$, $B= 0$ and taking
$\beta$ to depend only
 on dual coordinates, so that
\begin{equation}
R^{abc} =  3\partial^{[a} \beta^{bc]}\, . \label{Rsimple}
\end{equation}

An interesting observation from eq. (\ref{RFlux}) is that
any configuration defined by $\beta(y)$ leads to a
non-vanishing R-flux. Since locally $\beta$-transformations can be gauged
away, one could hope to construct locally geometric backgrounds
associated to the R-flux.

\end{itemize}
The $O(d,d)$ covariance is responsible for the symmetry among
(\ref{Hsimple}), (\ref{omegaSS}), (\ref{omegaB})   and (\ref{Qbeta}),
(\ref{QSS}), (\ref{Rsimple}),  respectively.

Including extra vector fields as in subsection \ref{vectors}, one
can explicitly write the fluxes carrying indices ${M}=2d+N$. It is
interesting to see that setting $B=0, \beta=0$ one
recovers the electromagnetic fluxes. 
Indeed, one finds 
\bea 
{\cal F}^i_{ab}\equiv
F_{(2d+i)\tilde a\tilde b}&=&2(e^{-1})^{\alpha}{}_{a}(e^{-1})^\beta{}_b\partial_{[\alpha} V^i{}_{\beta]}\, . 
\eea

Having made contact with string fluxes, we now define the subgroup
of $O(d,d)$ transformations whose elements correspond to
T-dualities, namely
\begin{equation}
(T_\alpha)^{ N}{}_M = \delta^{  N}{}_M - \delta^{N,
\alpha}\delta_{M, \alpha} -  \delta^{N, \alpha+d}\delta_{M,
\alpha+d} +  \delta^{N, \alpha+d}\delta_{M, \alpha} +
\delta^{N,\alpha}\delta_{M, \alpha+d} \, ,\label{TDualityTransf}
\end{equation}
where the index $\alpha$ denotes the direction in which the
T-duality is performed. As expected, these elements satisfy the
following properties
\begin{equation}
\left[T_\alpha,\ T_\beta \right] = 0\ , \ \ \ \
({T_\alpha})^N{}_P({T_\alpha})^{P}{}_M
 = \delta^N{}_M\, .
\end{equation}
The T-duality operators  (\ref{TDualityTransf})
lead to the Buscher rules introduced in \cite{Buscher:1987sk}. We have
seen in (\ref{ODDinv}) how the  generalized metric transforms under these
symmetries. Equivalently,
one can define a unified tensor  $E = g + B$ \cite{Giveon:1994fu}, which
transforms  as
\begin{equation}
U =\left(\begin{matrix}a & b  \\c  & d \end{matrix}\right) \ , \ \ \
\ E' = (a  E + b ) (c  E + d )^{-1}\, .
\end{equation}
Taking the $O(d,d)$ transformation to be a T-duality defined in
(\ref{TDualityTransf}) in the $y^\alpha$ direction, one obtains the
following transformations for $E$
\begin{equation}
E_{\alpha\alpha}' = \frac{1}{E_{\alpha\alpha}}\ , \ \ \ \ E'_{\beta
\alpha} = \frac{E_{\beta \alpha}}{E_{\alpha\alpha}} \ , \ \ \ \
E'_{\alpha \beta} = -\frac{E_{\alpha\beta}}{E_{\alpha\alpha}} \ , \
\ \ \ E'_{\beta\gamma} = E_{\beta\gamma} - \frac{E_{\beta\alpha}
E_{\alpha\gamma}}{E_{\alpha\alpha}}\, ,
\end{equation}
which  decomposed into symmetric and anti-symmetric parts lead to
the well-known Buscher rules
\begin{equation}
g'_{\alpha\alpha} = \frac{1}{g_{\alpha\alpha}} \ , \ \ \ \
g'_{\alpha \beta} = - \frac{B_{\alpha \beta }}{g_{\alpha\alpha}}\ ,
\ \ \ \ g'_{\beta \gamma} = g_{\beta \gamma} - \frac{g_{\alpha
\beta} g_{\alpha \gamma} - B_{\alpha \beta}B_{\alpha
\gamma}}{g_{\alpha\alpha}} \, ,\nn
\end{equation}
\begin{equation}
B'_{\alpha \beta} = - \frac{g_{\alpha \beta }}{g_{\alpha\alpha}}\ ,
\ \ \ \ B'_{\beta \gamma} = B_{\beta \gamma} - \frac{g_{\alpha
\beta} B_{\alpha \gamma} - B_{\alpha \beta}g_{\alpha
\gamma}}{g_{\alpha\alpha}} \, .\label{Buscher}
\end{equation}
Defined like this, the standard T-duality chain
\cite{Shelton:2005cf} holds
\begin{equation}
H_{abc}\ \ {\stackrel{T_a} {\longleftrightarrow}}\ \ \omega^a_{bc}\
\ {\stackrel{T_b} {\longleftrightarrow}}\ \ Q^{ab}_c\ \
{\stackrel{T_c} {\longleftrightarrow}}\ \ R^{abc}\, ,
\end{equation}
and when the Jacobi identities (\ref{JacobiFlux}) are written in terms of
the fluxes (\ref{FluxDefhet}), the following results of
\cite{Shelton:2005cf} are exactly recovered
\begin{eqnarray}
H_{e[ab}\omega^e_{cd]} &=&0\, ,\label{STWconst}\\
\omega^a_{b[c}\omega_{de]}^b + H_{b[cd}Q_{e]}^{ab} &=& 0\, ,\\
Q^{[ab]}_c \omega^c_{[de]} - 4 \omega^{[a}_{c[d} Q^{b]c}_{e]} +
H_{c[de]} R^{[ab]c} & = & 0\, ,\\
Q^{[ab}_c Q^{d]c}_e + \omega^{[a}_{ce} R^{bd]c} &=& 0\, ,\\
Q^{[ab}_c R^{de]c}&=&0\, .
\end{eqnarray}
This set of equations closes under  T-duality transformations
(\ref{TDualityTransf}), as expected from the fact that
(\ref{JacobiFlux}) is $O(d,d)$ covariant. Note that defining
${\cal E}_a = - Z_a$ and ${\cal E}^a = - X^a$ the T-covariant
algebra of \cite{Shelton:2005cf} is also recovered
\begin{eqnarray}
\left[Z_a,\ Z_b \right] &=& H_{abc}X^c+ \omega^c_{ab} Z_c \, ,\\
\left[Z_a,\ X^b \right] &=& -\omega^b_{ac} X^c+ Q^{bc}_{a} Z_c \, ,\\
\left[X^a,\ X^b \right] &=& Q^{ab}_{c}X^c + R^{abc} Z_c\, .
\end{eqnarray}

Interestingly enough, also the extra gauge  vectors fit in this
description. This can be seen by  using the Jacobi identity
(\ref{JacobiFlux})  with generic fluxes including indices $i=1\dots
N$ for vectors. Indeed, by keeping (for simplicity) only
$\omega_{ab}^c,\ H_{abc}$ and ${\cal F}_{ab}^i$ fluxes we obtain
\begin{eqnarray}
{\omega}_{[ab}^d {\omega}_{c]d}^e& = &0\ ,\\
  {\omega}_{[ab}^d  {\cal F}^i_{c]d} & = &0\ ,\label{fww}\\
  {H}_{e[ab}   {\omega}_ {cd]}^e&= &-
{\cal F}^i_{[ab}{\cal
  F}^i_{cd]}\ .\label{htad}
\end{eqnarray}
which are the Jacobi identities derived from the   heterotic algebra found in
\cite{Kaloper:1999yr}, after adjusting some normalization factors.

\subsection{Type I interpretation}

As we mentioned at the beginning of this section, the effective DFT
four dimensional theory can be equally well interpreted in terms of
a Type IIB orientifold compactification. For instance, if instead of
the identifications (\ref{FluxDefhet}) we had used
\begin{equation}
F_{\tilde a \tilde b \tilde c} = {\mathbf{F}}_{abc} \ , \ \ \ \
F^{\tilde a}_{\ \ \tilde b \tilde c} = \omega^{ a}_{ bc}\ , \ \ \ \
 F^{(2d+i)}{}_{\tilde a\tilde b} = {\cal F}^i_{ab} \, . \label{FluxDefIIB}
\end{equation}
where now ${\mathbf{F}}_{abc}$ is the RR 3-form flux of Type IIB
compactification with O9 planes, $\omega^{a}_{bc}$ is a geometric flux
and for simplicity we turn off the non-geometric fluxes. In this
case, the Jacobi  identities would read
\begin{eqnarray}
{\omega}_{[ab}^d {\omega}_{c]d}^e& = &0\ ,\\
  {\omega}_{[ab}^d  {\cal F}^i_{c]d} & = &0\ ,\\
  {\mathbf{F}}_{e[ab}   {\omega}_ {cd]}^e &= &-
{\cal F}^i_{[ab}{\cal
  F}^i_{cd]}\ ,\label{tad}
\end{eqnarray}
where the first equation is the usual constraint on geometric
fluxes, the second equation is the requirement of absence of
Freed-Witten anomalies and the third one is a tadpole cancelation equation
\cite{Aldazabal:2008zza} if vector fluxes are now interpreted as D9
magnetic fluxes.

The full set of Type IIB generalized fluxes can be identified (see for
instance \cite{Aldazabal:2008zza} in an O3 notation). For bulk fluxes, they
correspond to a definite    $SL(2,\mathbb{R})$ spin (say plus
in the notation of Section 4) in the  $\bf (2, 220) $ representation of
${SL(2,\mathbb{R})\times O(6,6)}_B$. The opposite spin would correspond
to S-dual fluxes.

From the point of view of the starting DFT, the original fields have
now a different interpretation. Namely, the ${\hat B}_{\hat \rho\hat
\nu}$ field entering the generalized metric $\hat{\cal H}_{\hat
M\hat N}$ in (\ref{MetricaGen}) is such that in the ten dimensional
effective field theory it reproduces the two form RR field. Namely,
the DFT effective theory in ten space-time dimensions now reads
\begin{eqnarray}
S_{10}&=&\int d^{10}x \sqrt{|g(x)|}e^{-2\Phi(x)} \left\{ {\bf R} ~+ ~
4 ~\partial ^\mu \Phi\partial_\mu \Phi -
\frac{1}{12}  { F}^2\right\}\, ,\nn
\end{eqnarray}
where  $ F$ is the RR 3-form field strength.

\section{Conclusions}\label{conclu}

We have performed a generalized Scherk-Schwarz compactification of
DFT and obtained the dimensionally reduced effective action in
arbitrary number of dimensions. The reduced action has  $O(d,d)$ 
symmetry broken by the fluxes 
and a gauge symmetry, both inherited from the global
$O(D,D)$ and gauge invariance of DFT, respectively. The structure
constants of the gauge group are defined by the C-bracket and we have given
their explicit definition in terms of the twists. We have
shown that, although the C-bracket does not generically satisfy the
Jacobi identity, it leads to gaugings which verify such identity when
particularly evaluated in terms of the twists.

Specified to four space-time  and six internal
dimensions, the effective action  reproduces   the  bosonic electric
sector of gauged
${\cal N}=4$ supergravity. The comparison was made with the
formulation of \cite{Schon:2006kz}, and the precise
dualizations required to make the equivalence explicit can be found
in Section 4.
Furthermore,
the global symmetry group of the
effective action was enhanced  to $O(d,d+N)$, 
by considering, as a starting point,
the abelian heterotic extension of DFT formulated in
\cite{Hohm:2011ex}. In addition, the gauging of
a generalization of the rescaling
symmetry considered in \cite{Derendinger:2007xp} was proposed as an origin
for the $\xi_+^M$ fluxes.
Altogether, our
results imply that all the electric gaugings of ${\cal N}=4$
gauged supergravity can be reached from DFT.

It is worth mentioning that throughout the paper  a strong constraint, as
discussed in section 2, was assumed. This is   needed  for gauge invariance of
the DFT original action.
Interestingly enough   the Scherk-Schwarz like dimensional reduction  to go from
DFT action (\ref{action}) to the effective action (\ref{our}) does not
require further use of the strong constraint\footnote{In the $\xi_{+M}\ne 0$
case, level matching is not enough and extra requirements, but still weaker than
the strong constraint, are needed.}. This also
applies, in particular,
for Jacobi identities  implying  the ``${\cal N}=4$ quadratic constraints", needed in
gauged supergravities for the closure of the gauge transformations.   
A similar observation was pointed out in  \cite{Geissbuhler:2011mx} (see also
\cite{Hohm:2011cp}).  It would then be interesting to analyze in
the future if this is an indication of some possible consistent relaxation of
the strong constraint of the starting DFT.

The generalized Scherk-Schwarz reduction can be interpreted as a
compactification on a twisted double  torus, its vielbein being
defined by the twist matrix. A precise definition of torsion
in terms of the connection allows to relate these geometric
concepts with the
fluxes. 
Moreover, the standard NS-NS (non)-geometric fluxes
($H, \omega, Q, R$) can be identified with
 the gaugings of the effective action, and 
the string $d$-dimensional background can be decoded from the double
twisted $2d$-torus. The fluxes obey the standard $T$-duality chain
and satisfy Jacobi identities reproducing the results of
\cite{Shelton:2005cf}. In this way, the
 higher dimensional origin of the string fluxes can be traced to the new
degrees of freedom of DFT.

Finally, we have presented a novel interpretation of  DFT in terms of Type
IIB/O9 orientifolds. Identifying the Kalb-Ramond
$B$-field with the RR 2-form $C_2$, the 10-dimensional action of the
Type I string can be obtained from DFT when dual coordinates are turned
off. This requires a reinterpretation of the
$O(D,D)$ symmetry, which is no longer trivially associated with (the
heterotic) T-duality. Also in this case the
gaugings can be identified with 
fluxes, and
some of the
quadratic constraints can be interpreted
in terms of tadpole cancelation conditions and
Freed-Witten anomalies.

\bigskip

{\bf \large Acknowledgments} { We are very grateful to   M. Gra\~na
for valuable comments on the manuscript. We also thank  A. Rosabal
and H. Triendl for useful discussions and comments. G. A., W.B. and C.N. 
thank IPhT- Saclay, where part of this work was done, for hospitality.
This work was supported by CONICET PIP 112-200801-00507, MINCYT
(Ministerio de Ciencia, Tecnolog\'\i a e Innovaci\' on Productiva of
Argentina), ECOS-Sud France binational collaboration project A08E06,
University of Buenos Aires UBACyT X161 and the ERC Starting
Independent Researcher Grant 259133 - ObservableString.}

\begin{appendix}
\section{Appendix}

\subsection{Gauging away $\beta$-transformations}\label{beta}

In this Appendix we closely follow the results of
\cite{Grana:2008yw}. The aim is to make
the structure of the vielbein 
explicit.
The internal vielbein ${\cal E}^A{}_{M}(x,\mathbb Y)$ 
which contains the fluctuations
around $U^A{}_M({\mathbb Y})$ (the 
vielbein of the background) is an element of
$G/K$, namely its planar index rotates under gauge transformations
  $K = O(d)\times O(d)$ from the left, and its curved index
rotates under global $G = O(d,d)$ transformations from the right.
Any two configurations that are connected by elements of $K$
physically correspond to the same configuration. Given that dim$(G)
= (2d-1)d$ and dim$(K) = d(d-1)/2 + d(d-1)/2$, then dim$(G/K) =
d^2$, of which $d(d+1)/2$ degrees of freedom are parameterized by a
metric field $h(x,{\mathbb Y})$, and the remaining $d(d-1)/2$ are
parameterized by an antisymmetric $b(x, {\mathbb Y})$-field.
A possible and convenient general parameterization of $\tilde {\cal E}$ is
given by
\begin{equation}
\tilde {\cal E} = \frac{1}{\sqrt{2}} \left(\begin{matrix}e_+ & -
e_+(b+{h})
\\ e_- & -e_-(b-{ h})\end{matrix}\right)\, , \label{VielbeinMasMenos}
\end{equation}
which, restricted to ${h}^{-1} = e_{\pm}^T e_{\pm}$, leads
to the usual form of the generalized metric
\begin{equation}
{\cal H}(x,\mathbb{Y}) = \tilde {\cal E}^T \tilde {\cal E}
 = \left(\begin{matrix}{h}^{-1} & -
{h}^{-1} b \\
b ~{h}^{-1} & {h} - b ~ {h}^{-1}
b\end{matrix}\right)\, .
\end{equation}

Given any element $\tilde {\cal O} \in K$ of the form
\begin{equation}
\tilde {\cal O} = \left(\begin{matrix}O_+ & 0 \\
0 & O_-\end{matrix}\right)\, ,
\end{equation}
acting on $\tilde {\cal E}$ from the left leaves the generalized metric
invariant, but rotates $e_\pm$. Notice that the vielbein
(\ref{VielbeinMasMenos}) is such that
\begin{equation}
{\cal I} = \tilde {\cal E}^T \left(\begin{matrix}-1&0\\0&
1\end{matrix}\right) \tilde {\cal E}\, ,
\end{equation}
so (\ref{VielbeinMasMenos}) is clearly not an element of $O(d,d)$.
However, it has the advantage that one knows how $O(d)\times O(d)$
transformations act on it. To take it to an $O(d,d)$ form preserving
the off-diagonal metric that we use in this paper, $\tilde {\cal E}$ should
be rotated by an  $SO(2d)$ transformation $R$,

\begin{equation}
R = \frac{1}{\sqrt{2}}\left(\begin{matrix}1 & 1
\\ -1 & 1\end{matrix}\right)\ , \ \ \ {\cal E} = R \tilde {\cal E} \ ,
\ \ \ {\cal I} = {\cal E}^T {\cal I} {\cal E}\, ,
\end{equation}
and this allows to determine the form of a generic $O(d)\times O(d)$
rotation in our case
\begin{equation}
{\cal O} = R \tilde {\cal O} R^T = \frac{1}{2} \left(
\begin{matrix}O_+ + O_-& O_- - O_+ \\ O_- - O_+& O_+ + O_-\end{matrix}\right)\,
.
\end{equation}

Note also that the form of the vielbein is now
\begin{equation}
{\cal E} = \frac{1}{2}\left(\begin{matrix} e_+ + e_- ~&~  
-(e_+ + e_-)b + (e_- -
e_+){h}\\
 e_- -e_+  ~&~ (e_+ -e_-)b + (e_+ + e_-) {h}
\end{matrix}\right) \, , \label{GeneralU}
\end{equation}
and we can perform a $K$ transformation with $O_\pm$ acting on
$e_{\pm}$ to set $e_+ = e_- = (e^{-1})^T$, and take ${\cal E}$ to
triangular form 
\begin{equation}\left(\begin{matrix}(e^{-1})^T & -(e^{-1})^T b\\ 0 & e
\end{matrix}\right) = \left(\begin{matrix}(e^{-1})^T & 0\\ 0 & e
\end{matrix}\right) \left(\begin{matrix}1& -b\\ 0 &
1\end{matrix}\right)\, .
\end{equation}
We have then managed to write it as a product of diffeomorphisms and
B-transformations. Any $\beta$-transformation
\be {\cal E}_\beta = \left(\begin{matrix}1 & 0\\ -\beta  &
1\end{matrix}\right)\ee
 would take it away from the triangular gauge, to a form
parameterizable as in (\ref{GeneralU}), so it can always be brought
to triangular form through a $K$-rotation. It is worth emphasizing
that, although the $K$-transformation seems to restore the effect of
the $\beta$-transformation, the restoration is only structural, and
in many cases it is performed at the expense of ending with
non-geometric backgrounds ${h}$ and $b$. In this paper we have chosen
the triangular gauge to do the computations, but  the results are
finally written in a covariant way.

Note that the gaugings are not invariant under $O(d)\times O(d)$ 
transformations of the vielbein, so even if it might seem that two physically
equivalent vielbeins give rise to different theories, the gaugings should
be computed through different expressions.

\subsection{A single flux example}

Here we would like to consider simple backgrounds, on which we apply
T-dualities and track how the obtained (non)geometries are encoded
in the double torus. For simplicity we  turn on a unique flux
for which  all the consistency equations are automatically
satisfied. We closely follow the procedure of
\cite{Dall'Agata:2007sr}, where it
is analyzed how the different fluxes twist the double torus
through successive T-dualizations and how
geometric and non-geometric backgrounds are obtained after
projecting the different twisted double tori to a base. Although
 the definition of flux is different here than in 
\cite{Dall'Agata:2007sr}, we show that the conclusions remain
unaltered.

\subsubsection*{$H$-flux}

The starting point is a double three-torus $T^3 \times \tilde T^3 =
(y^1, y^2, y^3, y^4, y^5, y^6)$ with H-flux over $T^3$  in Type IIB.
Although the base is $3$-dimensional, we are assuming that this
space can be fibered over some other space to render a proper
6-dimensional string background. We begin with a simple $T^3$ with
all radii equal to unity, described by
 ${U}=\mathbf{1}_6 $, and perform a $B$-transformation to
introduce $H_{123}=3N$ units of H-flux in $T^3$, namely
\begin{equation}
ds^2 = (dy^1)^2+(dy^2)^2+(dy^3)^2 \ , \ \ \ \ \ B = 6 N y^{[3}
dy^{1}\wedge dy^{2]} \, .\label{ToroConH}
\end{equation}
This leads to the twist matrix
\begin{equation}
U_H = \left(\begin{matrix}1&0&0&0&-N y^3&N y^2\\
0&1&0&N y^3&0&-N y^1 \\ 0&0&1&-N y^2& N y^1&0\\
0&0&0&1&0&0\\0&0&0&0&1&0 \\
0&0&0&0&0&1\end{matrix}\right)
\end{equation}
that allows to define a generalized coframe $\Gamma^A = U^{A}{}_P
d\mathbb{Y}^P$
\begin{eqnarray}
&& \Gamma^1 = dy^1 \ , \ \ \ \Gamma^2 = dy^2 \ , \ \ \ \Gamma^3 = dy^3 \,
,\label{OneForms}\\
&&\Gamma^4 = dy^4  - Ny^3  dy^2 +  Ny^2  dy^3 \, ,\cr&&  \Gamma^5 =
dy^5 + N y^3  dy^1 - N y^1 dy^3 \, ,\cr&&  \Gamma^6 = dy^6  - N y^2
dy^1 + N y^1 dy^2 \, .\nn
\end{eqnarray}
The identifications that make the one-forms (\ref{OneForms})
globally well defined are given by
\begin{eqnarray}
&& (y^1, y^5, y^6) \to (y^1 + 1, y^5 + N y^3, y^6 - N y^2) \, ,\ \ \   \ \ y^4
\to y^4  + 1\, ,\nn \\
&& (y^2, y^4, y^6) \to (y^2 + 1, y^4 - N y^3, y^6 + N y^1) \, ,\ \ \  \ \ y^5
\to y^5  + 1\, ,\nn \\
&& (y^3, y^4, y^5) \to (y^3 + 1, y^4 + N y^2, y^5 - N y^1) \, ,\ \ \  \
\ y^6 \to y^6  + 1 \, .\label{MonH}
\end{eqnarray}

One can define three kahler moduli as
\begin{equation}
\rho_i =  \frac 12\epsilon_{ijk}B_{jk} + i\ {\rm Vol} (T_{i}^2)
\label{Kahlers}
\end{equation}
and check that under the monodromies $y^i \to y^i + 1$ in
(\ref{MonH}) they shift as $\rho_i \to \rho_i  + N$.

\subsubsection*{$\omega$-flux}

Now we apply a T-duality in the direction $y^3$ leading to a IIA
background with flux $\omega^3_{12}= 3N$ where
\begin{equation}
U_\omega = \left(\begin{matrix}1&0& Ny^2&0&-N y^6&0\\
0&1&-N y^1& N y^6&0&0 \\ 0& 0&1&0&0&0\\
0&0&0&1&0& 0\\ 0&0&0&0&1&0 \\
0&0&0&-N y^2& N y^1&1\end{matrix}\right) \, .\label{NuOmega}
\end{equation}
The one forms
\begin{eqnarray}
&& \Gamma^1 = dy^1 \ , \ \ \ \Gamma^2 = dy^2 \ ,
\ \ \ \Gamma^6 = dy^6 \, ,\\ &&\Gamma^4 = dy^4  - N y^6 dy^2
+ N y^2 dy^6 \, ,\cr && \Gamma^5 = dy^5  +N y^6  dy^1- N y^1 dy^6\, ,\cr &&
\Gamma^3 =
dy^3 - N y^2 dy^1 + N y^1 dy^2\, ,\nn
\end{eqnarray}
are now globally well behaved under the shifts
\begin{eqnarray}
&& (y^1, y^5, y^3) \to (y^1 + 1, y^5 + N y^6, y^3 - N y^2) \, , \ \ \  \ \ y^4
\to y^4  + 1\, ,\nn \\
&& (y^2, y^4, y^3) \to (y^2 + 1, y^4 - N y^6, y^3 + N y^1) \, ,\ \ \   \ \ y^5
\to y^5  + 1\, ,\nn \\
&& (y^6, y^4, y^5) \to (y^6 + 1, y^4 + N y^2, y^5 - N y^1) \, ,\ \ \   \
\ y^3 \to y^3  + 1\, .
\end{eqnarray}

One can now read the new background from (\ref{NuOmega}).
After projecting to the base $y^6 = 0$, it is as expected a so-called
twisted torus
\begin{equation}
ds^2 = (dy^1)^2 + (dy^2)^2 + (dy^3 - N y^2 dy^1 + N y^1 dy^2)^2 \ ,
\ \ \ \ \ \ B = 0\, .\label{twistedtorus}
\end{equation}
This can also be obtained by applying a
T-duality (\ref{Buscher}) to (\ref{ToroConH}).
Note that when inserting (\ref{twistedtorus}) into (\ref{omegaSS}),
one obtains  $\omega^3_{12} = 2N$. The fact that the flux
is reduced by a unit of $N$ is a reflection of the projection to the
base.

 Acting with T-duality over (\ref{Kahlers}), the IIB kahler moduli $\rho_i$ are
maped into the complex structure moduli $\tau_i$ of the twisted
torus in IIA. In particular, the real part Re$(\tau_1) = - y_3 /
y_2$ shifts as $\tau_1 \to \tau_1 + N$ when $y^1 \to y^1 + 1$. Also,
under the shift $y^1\to y^1 + 1$, the ${\cal S}\in O(2,2)$
transformation needed to bring the metric to its original form is
block diagonal and does not mix the metric with the $B$-field, $i.e.$
\begin{equation}
{\cal S} = \left(\begin{matrix}S^{-  T}&0\\0& S\end{matrix}\right)\
, \ \ \ \ \  \ \ \ \ S =
\left(\begin{matrix}1&0&0\\0&1&0\\0&-N&1\end{matrix}\right)\ , \ \ \
\ \  \ \ \ \  U_\omega(y^1+1){\cal S} =  U_{\omega}(y^1) \, .
\end{equation}

\subsubsection*{$Q$-flux}

Now we apply a T-duality in the direction $y^2$ and obtain a
background with flux $Q^{23}_1 = 3N$ where
\begin{equation}
U_Q =  \left(\begin{matrix}1&-N y^6& Ny^5&0&0&0\\
0&1&0&0&0&0 \\ 0&0&1&0&0&0\\
0&0&0&1&0&0\\0&0&-Ny^1&N y^6&1&0 \\
0& Ny^1&0&-Ny^5&0&1\end{matrix}\right) \, ,\label{NuQ}
\end{equation}
with globally well defined one-forms on the double torus
\begin{eqnarray}
&& \Gamma^1 = dy^1\ , \ \ \ \Gamma^5 = dy^5 \ , \ \ \
\Gamma^6 = dy^6  \, ,\\ &&\Gamma^4 =
dy^4 - N y^6 dy^5 + N y^5 dy^6\, ,\cr&& \Gamma^2 = dy^2 +N y^6 dy^1-Ny^1 dy^6\,
,
\cr&& \Gamma^3 = dy^3 - N y^5 dy^1 + N y^1 dy^5 \, ,\nn
\end{eqnarray}
under the monodromies
\begin{eqnarray}
&& (y^1, y^2, y^3) \to (y^1 + 1, y^2 + N y^6, y^3 - N y^5) \, , \ \ \  \ \ y^4
\to y^4  + 1\, ,\nn \\
&& (y^5, y^4, y^3) \to (y^5 + 1, y^4 - N y^6, y^3 + N y^1) \, , \ \ \  \ \ y^2
\to y^2  + 1\, ,\nn \\
&& (y^6, y^4, y^2) \to (y^6 + 1, y^4 + N y^5, y^2 - N y^1) \, , \ \ \  \
\ y^3 \to y^3  + 1\, .
\end{eqnarray}

The matrix (\ref{NuQ}) does not have the proper triangular form to
read off the background. To take it to such form, we perform an
$O(3)\times O(3)$ rotation acting on the left, defined by
\begin{equation}
{\cal O} = \frac{1}{2}\left(\begin{matrix}O_+ + O_-& O_- - O_+\\
O_- - O_+ & O_+ + O_-\end{matrix}\right) \ , \ \ \ O_\alpha =
\left(\begin{matrix}1  & 0 & 0 \\ 0 &\gamma & -\alpha N y_1 \gamma
\\ 0 & \alpha Ny_1\gamma& \gamma\end{matrix}\right)\, ,
\end{equation}
with $\gamma \equiv 1/ \sqrt{1 + (N y^1)^2}$,
leading to
\begin{equation}
{\cal O}\ U_Q = \left(\begin{matrix}1& -N y^6 &  N y^5& 0& 0& 0
\\ 0& 1/\gamma& 0 & - N^2 y^1 y^5 \gamma & 0&   N y^1 \gamma \\ 0&0& 1/\gamma& -
N^2 y^1 y^6 \gamma& -N y^1 \gamma & 0 \\ 0& 0&0&1&0&0 \\ 0& 0&0& Ny^6 \gamma&
\gamma & 0 \\ 0&0&0&-N y^5 \gamma & 0& \gamma\end{matrix}\right)\,
.\end{equation}
We then see that the background is locally described by
\begin{equation} ds^2 = (dy^1)^2+ \frac{1}{1 + (N y^1)^2} ((dy^2)^2 + (dy^3)^2)
\ ,
\ \ \ \ \ B = -\frac{Ny^1}{1 + (N y^1)^2} dy^2 \wedge dy^3\,
,\label{Qbackground}
\end{equation}
which can  also be obtained by application of Buscher rules.
Again, if the fluxes are computed using this projected background,
they are reduced by one unit of $N$, i.e. $Q^{12}_3 = N$.

Now we can compute the IIB kahler modulus and its shift under $y^1
\to y^1 + 1$ as
\begin{equation}
\rho_1 = \frac{-1}{N y^1 + i} \ , \ \ \ \ \frac{1}{\rho_1} \to
\frac{1}{\rho_1} - N\, ,
\end{equation}
and this is again a fiber of $T^2_1$ shifting the kahler modulus as we
move in the direction $y^1$. Although this background is locally
described in terms of a metric and a B-field, upon going around a
cycle $y^1 \to y^1 +1$, the metric and B-field mix through an
${\cal S} \in O(2,2)$ transformation ${\cal O}U_Q (y^1+1){\cal S} = {\cal O}U_Q
(y^1)$,
 defined by
\begin{equation}
{\cal S} =\left(\begin{matrix}1 & 0&0&0&0&0\\ 0 &
\frac{\gamma(y^1+1)}{\gamma(y^1)} & 0 & 0 & 0 & -N
\gamma(y^1)\gamma(y^1+1)\\ 0&0& \frac{\gamma(y^1+1)}{\gamma(y^1)}&0 &
N\gamma(y^1+1)\gamma(y^1)&0
\\ 0&0&0&1&0&0 \\ 0&0&0&0& \frac{\gamma(y^1)}{\gamma(y^1+1)}&0\\ 0&0&0&0&0&
\frac{\gamma(y^1)}
{\gamma(y^1+1)}\end{matrix}\right)\, ,
\end{equation}
where we have already projected to the base. The fact that under a
cycle the metric and $B$-field mix implies that
this
background is not globally  well defined. However, as seen, locally it is
described by a metric and a $B$-field. Therefore, these backgrounds
are said to be locally geometric, but globally non-geometric.

\subsubsection*{$R$-flux}

Finally, we apply the remaining T-duality in the direction $y^1$ to
obtain a background with flux $R^{123} = 3N$ where
\begin{equation}
U_R =  \left(\begin{matrix}1&0&0&0&0&0\\
0&1&0&0&0&0 \\ 0&0&1&0&0&0\\
0&-N y^6& N y^5&1&0&0\\ Ny^6&0&-N y^4&0&1&0 \\
-Ny^5& N y^4&0&0&0&1\end{matrix}\right) \, .\label{NuR}
\end{equation}
This defines the one-forms
\begin{eqnarray}
&&\Gamma^4= dy^4 \ , \ \ \ \Gamma^5 = dy^5 \ , \ \ \ \Gamma^6  =
dy^6\, ,\\
&& \Gamma^1 = dy^1- N y^6  dy^5 +N y^5 dy^6\, ,\cr  && \Gamma^2 =
dy^2  +N  y^6 dy^4 - N y^4 dy^6\, ,\cr&& \Gamma^3 = dy^3 - N y^5 dy^4 + N y^4
dy^5 \, ,\nn
\end{eqnarray}
and the monodromies
\begin{eqnarray}
&& (y^4, y^2, y^3) \to (y^4 + 1, y^2 + N y^6, y^3 - N y^5)\, , \ \ \  \ \ y^1
\to y^1  + 1\, ,\nn \\
&& (y^5, y^1, y^3) \to (y^5 + 1, y^1 - N y^6, y^3 + N y^4)\, , \ \ \  \ \ y^2
\to y^2  + 1\, ,\nn \\
&& (y^6, y^1, y^2) \to (y^6 + 1, y^1 + N y^5, y^2 - N y^4)\, , \ \ \  \
\ y^3 \to y^3  + 1\, .
\end{eqnarray}

The matrix $U_R$ reduces to the identity when projected to the base,
and must then be associated to a trivial torus with vanishing
$B$-field. This can be understood if we take into account that in
every step of the T-duality procedure we project out the new dual
coordinates, and this successively reduces the flux by  units of
$N$. Since we started from a background with  $H_{123} = 3N$ and
performed three T-dualities, we have ended after three successive
projections with a background having $R^{123} = 0$. In summary,
T-dualities incorporate dual fluxes into the game which we are
forced to project out if we want to make contact
with some string background geometry. In this example, this can be done
for $H$, $\omega$ and $Q$ but not for $R$ which then receives the
name of locally non-geometric flux.

\end{appendix}

\end{document}